\documentclass[10pt,journal,compsoc]{IEEEtran}
\ifCLASSOPTIONcompsoc
  \usepackage[nocompress]{cite}
\else
  \usepackage{cite}
\fi
\ifCLASSINFOpdf
\else
\fi

\usepackage[utf8]{inputenc} 
\usepackage[T1]{fontenc}    
\usepackage{url}            
\usepackage{booktabs}       
\usepackage{amsfonts}       
\usepackage{nicefrac}       
\usepackage{microtype}      

\usepackage{caption}
\usepackage{subcaption}

\usepackage{amsmath, amssymb, amsthm, enumerate, graphicx}
\usepackage[usenames,dvipsnames]{color}
\usepackage{bm}
\usepackage[colorlinks=true,urlcolor=blue]{hyperref}
\usepackage[utf8]{inputenc}
\usepackage{amssymb}
\usepackage{setspace}
\usepackage{marvosym}
\usepackage{wrapfig}
\usepackage{datetime}
\usepackage[many]{tcolorbox}
\usepackage{array}
\usepackage{multirow}
\usepackage{wasysym}
\usepackage{cancel}
\usepackage{fancyhdr}
\usepackage{listings}
\usepackage{color}
\usepackage{caption}
\theoremstyle{definition}
\usepackage{cite}

\newtheorem{theorem}{Theorem}[section]

\newtheorem{definition}{Definition}[section]

\begin{document}
%
\title{Dynamic Coupling Strategy for Interdependent Network Systems Against Cascading Failures}
%
%
%
%

\author{I-Cheng Lin, Osman Ya\u{g}an, Carlee Joe-Wong
\IEEEcompsocitemizethanks{\IEEEcompsocthanksitem I-Cheng Lin is with the Department of Electrical and Comnputer Engineering, Carnegie Mellon University, Pittsburgh, PA 15213.\protect\\
E-mail: ichengl@andrew.cmu.edu
\IEEEcompsocthanksitem Osman Ya\u{g}an is with the Department of Electrical and Computer Engineering, Carnegie Mellon University, Pittsburgh, PA 15213.\protect\\
E-mail: oyagan@andrew.cmu.edu
\IEEEcompsocthanksitem Carlee Joe-Wong is with the Department of Electrical and Computer Engineering, Carnegie Mellon University, Pittsburgh, PA 15213.\protect\\
E-mail: cjoewong@andrew.cmu.edu
}
\thanks{}}

%
%

\markboth{}%
{}
%



\IEEEtitleabstractindextext{%
\begin{abstract}
Cascading failures are a common phenomenon in complex networked systems where failures at only a few nodes may trigger a process of sequential failure. We applied a flow redistribution model to investigate the robustness against cascading failures in modern systems carrying flows/loads (i.e. power grid, transportation system, etc.) that contain multiple interdependent networks. In such a system, the coupling coefficients between networks, which determine how much flows/loads are redistributed between networks, are a key factor determining the robustness to cascading failures. We derive recursive expressions to characterize the evolution of such a system under dynamic network coupling. Using these expressions, we enhance the robustness of interdependent network systems by dynamically adjusting the coupling coefficients based on current system situations, minimizing the subsequent failures. The analytical and simulation results show a significant improvement in robustness compared to prior work, which considers only fixed coupling coefficients. Our proposed Step-wise Optimization (SWO) method not only shows good performance against cascading failures, but also offers better computational complexity, scalability to multiple networks, and flexibility to different attack types. We show in simulation that SWO provides robustness against cascading failures for multiple different network topologies. 
\end{abstract}

}

\maketitle

\IEEEdisplaynontitleabstractindextext

%
\IEEEpeerreviewmaketitle

\IEEEraisesectionheading{\section{Introduction}\label{sec:introduction}}

%
%
%
%
Large-scale networked systems, such as the Internet of Things (IoT), the urban transportation network, the smart power grid and other national infrastructures,  have become an integral part of our daily lives. It has been observed \cite{rassia2017smart,10.1093/nsr/nwu020} that these large-scale networks do not work in isolation, but instead are highly interdependent with each other. For example, urban transportation systems consist of general road networks, bus systems, subway systems, train / high speed rail and other railway systems and bike sharing systems. These networks are interdependent in the sense that passengers may switch between them.

Much research on the robustness of such network systems focuses on the phenomenon called cascading failures. Cascading failures are a common phenomenon in complex networked systems where the failure of a small number of nodes may trigger a process of sequential failure, eventually making the whole system break down. The blackout in Italy on September 28, 2003 \cite{buldyrev2010catastrophic} is an example of this process where the failure of a few power lines eventually caused a serious, nation-wide power outage. Such cascading failure processes can also take down other network systems like railway networks, communication networks etc. \cite{buldyrev2010catastrophic}. In this work, we focus on load-carrying, interdependent networks that can offload to each other.

Existing works use different models to study the dynamics of this process. Research on network robustness against cascading failures considered a single network case \cite{dobson2003probabilistic, dobson2005loading, crucitti2004model, motter2002cascade, xia2010cascading, yaugan2015robustness, zhang2016optimizing, scala2016cascades} with different models that could be applied to different kind of networks. We focus in this work on networks that carry resource flows, such as electric grids or transportation networks. Prior models of these networks \cite{yaugan2015robustness, zhang2016optimizing}  can be optimized against cascading failures. However, as stated above, most modern systems exhibit interdependencies with each other, which prior works for load-carrying networks have not fully considered. The robustness of interdependent network systems were studied in \cite{gao2012networks, 1265180, parshani2010interdependent, 6148227, PhysRevE.83.065101, duan2019universal, buldyrev2010catastrophic, brummitt2012suppressing, bianconi2014multiple, di2016cascading} with different models of networks \cite{cheng2015cascade, PhysRevLett.108.228702} used to investigate different real-world network systems \cite{huang2015cascading, tang2016complex}. Past research on percolation-based models \cite{buldyrev2010catastrophic, buldyrev2011interdependent, gao2012networks, son2012percolation} shows that interdependent networks could be more vulnerable to cascading failures than an isolated network \cite{vespignani2010fragility}. A node in such model is assumed to be functioning if it belongs to the giant component (i.e., connected to the rest of the nodes) and failed otherwise. Such a model can capture the operation of systems like communication networks which focus on the connectivity between nodes. However, the percolation model cannot capture the characteristics of load-bearing systems such as the power grid or transportation systems. In such systems, the failing of a node in the network may result in a redistribution of the load or flow it carried to other functioning nodes. The extra load being redistributed may cause further failure in the nodes that receive load that exceeds their capacity. One special case is that of infrastructure networks \cite{o2007critical, dobson2007complex} and the cyber-physical systems based on physical infrastructures \cite{zhang2019robustness}, where the cyber-network represents the communication network, and the physical-network representing the real-world, flow-carrying network. The cyber-network can be modeled with a percolation-based model, while nodes in the physical-network function only if the flow they carry is below their capacities. In this work, we instead focus on interdependent networks with homogeneous flow/load, where the flow/load could be transferred between the networks. 

We focus on systems where different networks in the system carry the same type of load/flow (e.g., passenger flow in different transportation networks or power load in different layers of the power grid). To capture the characteristics of systems carrying flows or load, a flow/load redistribution model is applied to investigate the robustness of such interdependent systems. The coupling coefficients between networks have been shown to be a key factor determining the robustness of such systems \cite{zhang2018cascading}. Based on the choice of coupling coefficient between networks, the robustness of the system could be maximized at a non-trivial coupling coefficient in general. Therefore, the interdependence between networks could lead to an improved robustness with a proper choice of coupling coefficient. 

While prior works have shown the importance of choosing the right coupling coefficients, they have not fully investigated how to do so. We solve this problem and further propose the first (to the best of our knowledge) dynamic strategy to determine the right coupling coefficients in each step of the cascading failure process. We study the case with random initial failure and use the flow/load redistribution model, which has been widely used to study the cascading failure process in several kinds of systems \cite{yaugan2015robustness, zhang2018cascading}.  Prior works have studied fixed coupling coefficients~\cite{zhang2018cascading}. For a multiple network system, however, dynamically adjusting coupling coefficients between networks helps improve the system's robustness against the cascading failure phenomenon. For example in transportation network, with the integration of the modern intelligent transportation system, if the network is impacted by certain disaster, we could guide the extra traffic flow to other interdependent networks. That is, if a metro line's service is reduced, passengers can be redirected to buses and taxis. At every stage of the failure process, we could determine the amount of the customers to be guided towards using different transportation system. A key challenge is then to determine the coupling coefficients (i.e., the fraction of users to be guided between different networks) that will increase robustness to cascading failures. Adding to this challenge, the optimal coefficients may change significantly during the cascading failure process. 
For example, if the initial failure occurred in the subway system, we may want to guide more passengers to use other systems to prevent the sudden congestion of the subway network. After the ratio of passengers using different networks changes, we may want to reach a new equilibrium state to avoid overflowing traffic flow to other networks, making the utilization of the subway system become too low. Thus, dynamically adjusting the setting of coupling coefficients based on current system situation becomes another key challenge.

We summarize the above discussion into two critical problems to solve. First, how do we \emph{characterize the system} (i.e., the average capacity remaining, total extra flows) under the case with dynamic coupling coefficients between networks at each time step? To do so, we should predict not just whether the system will ultimately survive or not, but also its final system status (e.g., the fraction of surviving nodes and free space left to accommodate more failures). Making such a prediction is difficult since in each step, dynamically adjusting the coupling coefficients may cause the system status varies a lot in the following steps. Second, is there any \emph{strategy to determine the coupling coefficients at each time} to guarantee a better robustness of the system against cascading failure? As stated above, during the cascading failure process, the choice of coupling coefficients may significantly impact the subsequent system dynamics, and characterizing this future impact is nontrivial. For instance, if the initial failure starts from the system with much higher capacity like subway system, guiding more passengers to use other lower capacity systems like taxis may slow down the propagation of failure in the subway system, but potentially causes the whole taxi network to be congested. The passengers using taxis may need to use the subway system, causing more pressure on the subway system than before. Moreover, if there are constraints on the coupling coefficients, determining the optimal coupling coefficients becomes still more challenges. Take the transportation system as an example: we may not be able to guide all excess passengers currently using subway system to switch to another system, due to the physical constraints between different transportation systems.

\textbf{Our contributions} in this work are the following: 
\begin{itemize}
    \item We \emph{characterize the final system size} of a system with multiple interdependent networks under time-varying coupling coefficients. We are interested in the final system size instead of other metrics like the overall free space of the surviving nodes since it directly indicates the portion that the system is still functioning. For example, in an urban transportation system, we would care about how many areas of the city aren't being congested but not how much extra traffic the system could accommodates after the cascading failure.
    \item We \emph{create a framework to dynamically adjust the coupling coefficients between networks} based on the expected system state at each time, minimizing the subsequent extra load. Our strategy aims to further optimize the robustness of interdependent network systems. Compared to searching for the optimal fixed coupling coefficients, our algorithm solves an optimization problem at each time step, which takes much less time as the optimal fixed coupling coefficients could only be found by brute force search in recent works \cite{zhang2018cascading}.
    \item We \emph{empirically show that our dynamic strategy have the same performance compare to optimal fixed coupling coefficients strategy} and, unlike a fixed strategy, can automatically adapt to handle different kinds of attacks. Our proposed dynamic strategy further outperforms other dynamic heuristics.
    \item We \emph{provide additional simulations showing that our proposed strategy continues to perform well when the network is not fully-connected}. 
    Although our theoretical analysis assumes fully connected networks, these simulations provide more insight in our strategy's performance on more realistic network topologies, such as transportation networks (where the network topology may be based on geographical locations) or power networks (where stations/transmission lines near the generator or the customers may have different node degrees). According to other previous works \cite{zhang2019robustness}, not all kinds of the networks having the same capacity and initial load could have the same robustness. We analyze how our proposed method as well as the fixed coupling coefficients strategy perform under different network topologies. 
    \\
\end{itemize}

The rest of the paper is organized as follows. We describe the details of the system model in Section \ref{sec:model} and define the parameters that will been used throughout the whole paper. Inspired by the work \cite{zhang2018cascading}, we present the mathematical analysis of such an interdependent system in Section \ref{sec:math}. In addition, we give a clear definition and reasoning of our dynamic coupling strategy while providing the detailed analysis of the special case with uniform free space distribution to better illustrate how our strategy works. In Section \ref{sec:num}, we provide the numerical and simulation results. We apply our strategy to both identical networks and non-identical networks settings and compare with other strategies including the fixed coupling coefficient strategy. We have a discussion in Section \ref{sec:discussion} and then conclude our work in Section \ref{sec:conclusion}.

\section{Model Description}\label{sec:model}

\subsection{Network Model}
The interdependent network system consists of multiple networks. Let $\mathcal{N}$ denote the set of networks in the system, where $|\mathcal{N}|=N$. Each network $\mathcal{X} \in \mathcal{N}$ is a set of nodes and edges between nodes. There are $N_{\mathcal{X}}$ nodes. We follow the widely used \cite{daniels1945statistical, andersen1997tricritical, pahwa2014abruptness, yaugan2015robustness} fiber-bundle model, which has been used to investigate the breakdown of a wide class of systems \cite{zhang2016optimizing, zhang2018cascading, moreno2001model, turcotte2004damage}. We assume that each node inside the same network is fully-connected, i.e., when a node fails, its load will be equally redistributed to all other nodes in the network. We relax this assumption in the last part of Section \ref{sec:num}'s simulations. We assume each node carries a given amount of load, for example representing a transmission line in a power network or a road segment in a transportation network. The $i$-th node in network $\mathcal{X}$ has three different characteristics, which are \emph{load} $L_{\mathcal{X}i}$, \emph{free space} $S_{\mathcal{X}i}$ and \emph{capacity} $C_{\mathcal{X}i}$. The load indicates the current load that the node has (i.e., the power load carried by a transmission line or a traffic load passed through a road segment), and the capacity of a node indicates the maximum load it could accommodate (i.e. the maximum power that a transmission line could carry or the maximum traffic that a road segment could handle). The free space indicates the space to accommodate extra loads, which is the difference between the capacity and current load. Thus, the three characteristics have the relation that $C_{\mathcal{X}i} = L_{\mathcal{X}i} + S_{\mathcal{X}i} \: \forall \, i \in \mathcal{X}, \: \mathcal{X} \in \mathcal{N}$.

\subsection{Initial Conditions}
For each network $\mathcal{X} \in \mathcal{N}$, the initial load and initial free space of the nodes in the network follow certain given probability distributions $f_{L, \mathcal{X}}$ and $f_{S, \mathcal{X}}$. The initial load and initial free space are non-negative, that is:
\begin{equation}
L_{\mathcal{X}i} \geq 0, \; \\
S_{\mathcal{X}i} \geq 0, \; \\
\forall i \in \mathcal{X}, \mathcal{X} \in \mathcal{N} \\
\end{equation}
Also, the initial load and initial free space are not correlated, following past works \cite{zhang2018cascading}, and their distributions for different networks need not be the same.

Initially, each network receives a random initial attack. This captures the failure of nodes in the network like the failure of transmission lines in a power network or failure / congestion of road segments in a transportation network caused by either a natural or man-made disaster. The attack size of network $\mathcal{X} \in \mathcal{N}$ is denoted as $p_{\mathcal{X}}$ where $p_{\mathcal{X}} \in [0, 1]$, which means that initially $p_{\mathcal{X}}$ fraction of nodes in network $\mathcal{X}$ fail. These failures initiate the cascading failure process, as the initial loads of the nodes being attacked will now be redistributed to other nodes within the system. Nodes that are not attacked initially will not be directly harmed, i.e., they will not fail unless they receive more load than they could handle during the cascading process, as we explain in the next section.
We list the initial parameters defining the system in Table~1.

\begin{table}[hbt]
\begin{tabular}{ |p{1.2cm}|p{6.0cm}|  }
 \hline
 \multicolumn{2}{|c|}{List of initial parameters} \\
 \hline
 Notation & Definition \\
 \hline
$p_A, p_B$ & Initial fraction of failure (or failure probability) of network A and B.\\
 \hline
$N_A, N_B$ & Initial number of nodes of network A and B. \\
 \hline
$L_A, L_B$ & Mean of initial load in network A and B. \\
 \hline
$S_A, S_B$ & Free space of network A and B, each follow certain probability distribution. \\
 \hline
\end{tabular}
\captionof{table}{Initial Parameters of the System}\label{IniParaDef}
\end{table}

\subsection{Cascading Failure Process and the Load Redistribution Policy}

The cascading failure process of the network system we described above is triggered by an initial attack. The attack causes the nodes to fail and once a node fails, its load will be redistributed to other nodes. For example, once a transmission line in a power system fails, to keep the whole system operating, the original power load carried by the failed line will be redistributed to other transmission lines. Nodes may also receive loads from failed nodes in other networks, the amount of which is determined by our chosen coupling coefficients, which we will explain in detail later. If the extra load received by a surviving node exceeds its current free space, the node will fail. In other words, for the $i$-th node in network $\mathcal{X}$ to fail at time step $t$, (i) the total load it carries before receiving the extra load $L_{\mathcal{X}_i}(t-1)$ should be less than its capacity $C_{\mathcal{X}_i}$, i.e., $L_{\mathcal{X}_i}(t-1) \leq C_{\mathcal{X}_i}$ and (ii) the load it carries after receiving the extra load redistributed at time $t$ should exceed its capacity, i.e., $L_{\mathcal{X}_i}(t) > C_{\mathcal{X}_i}$. If the node fails at time $t$, its load will be redistributed to other neighboring surviving nodes as well as the nodes in other interdependent networks at the next time step $t+1$. 
This process continues, causing a cascade of failures. Furthermore, this process could not be reversed. That is, once a node failed, it could not be recovered. The cascading failure process will eventually be halted after meeting one of two different stopping conditions:

\begin{itemize}
    \item \textbf{Surviving Case:} At a certain step, the remaining nodes absorb the extra load and do not cause any more failures, stopping the cascading failure process. In this situation, a certain portion of nodes remain alive, which we call the “surviving portion”. 
    \item \textbf{Breakdown Case:} After some number of steps, there is no node which has survived in the system. The extra load has nowhere to be redistributed, and the whole system breaks down.
    \\
\end{itemize}
 The relation between the initial attack size and the final system size is not a smooth curve. In fact, when the initial attack size exceeds a certain threshold, the final size of the whole system would have a transition between two phases. Referring to Figure \ref{fig:HET} and Figure \ref{fig:SCTN}, we can see all curves have a sudden drop when we try to increase the initial attack size, separating the two phases. In a single network system, there would only be one transition, which is the transition between the first and the second situation listed above. This transition is being proved in \cite{zhang2018cascading} for fixed coupling coefficients cases. We define the \emph{critical attack size} as our metric for the network robustness, which specifies the minimal initial attack size that could cause the whole system to break down in the end. That is, for any initial attack size greater than the critical attack size, the system could not survive. 

\textbf{Load Redistribution Policy.} In a system with multiple interdependent networks, the extra load could be redistributed in the same network or redistributed to other networks. The load being redistributed inside the network captures the nature of the load being shared internally. Many network systems in the real world, such as the power grid \cite{brummitt2012suppressing} or transportation networks, shed their extra load to other interdependent networks. Our model's redistribution of loads to other networks captures this phenomenon. For a $n$-network system, we define the \textbf{coupling matrix} at time $t$ as $\mathcal{M} (t)$ with size $n \times n$, where the element $m_{i, j} (t)$ at the $i$-th row, $j$-th column of the matrix is the fraction of the extra load to be redistributed from network $i$ to network $j$. In other words, this specifies the coupling coefficient from the $i$-th network to the $j$-th network. The diagonal elements of the matrix indicate the portion of the extra load of each network to be redistributed inside itself. Since the extra load will not disappear, the sum of the portion of the extra load to be redistributed in every network should be 1. That is:

\begin{equation}
\sum_{i=1}^{n} m_{i, j} (t) = 1 \; \forall j \in \mathbb{N}, \: 1 \leq j \leq n \\
\end{equation}

We define the sum of the total extra load in network $i$, at time $t$ to be redistributed as $F_{i} (t)$. Then the extra load from the $i$-th network to be redistributed in the current network will be $ F_{i} (t) \cdot m_{i, i} (t)$, and the extra load to be redistributed to the $k$-th network where $k \in \mathbb{N}, \: 1 \leq k \leq n, \: k \neq i$ is $ F_{i} (t) \cdot m_{i, k} (t)$.

Thus, if we define the total extra load to be redistributed from the $i$-th network at time $t$ where $1 \leq i \leq n$ to be $F_{i} (t)$, then $R_{k} (t)$, the total extra load the $k$-th network received at time $t$, is:

\begin{equation}
R_{k} (t) = \sum_{i=1}^{n} F_{i}(t) \cdot m_{i, k} (t) \; \forall k \in \mathbb{N}, \: 1 \leq k \leq n \\
\end{equation}

In this work, we use the equal redistribution model. Though simple, it captures the nature of failure propagation in physical systems and lets us focus on the phenomenon of interdependence between networks. Under the equal redistribution model, if the number of surviving nodes in the $i$-th network at time $t$ is $N_{i} (t)$, then $\Delta L_{i} (t)$, the extra load received by a single node in the $i$-th network at time $t$ will be:

\begin{equation}
\Delta L_{i} (t) = \frac{R_{i} (t)}{N_{i}(t)} \\
\end{equation}

Thus, for a single node $j$ in the $i$-th network, the load update will be:

\begin{equation}
L_{i, j}(t) = L_{i, j} (t-1) + \frac{R_{i} (t)}{N_{i}(t)} \\
\end{equation}

At time $t$, the condition for node $j$ to fail will be: $ L_{i, j}(t) > C_{i, j}$, which is the mathematical expression representing the total load on node $j$ exceed its capacity.

\subsection{The Dynamic Coupling Coefficient Strategy}

The coupling coefficients that specify the portion of extra flow to be redistributed across different networks significantly affect the robustness of our system, which we quantify with the critical attack size. A fixed coupling coefficients strategy with flow-redistribution based model of interdependent system was studied in past works~\cite{zhang2018cascading}. It shows that under the same system settings, varying the coupling coefficients can change the critical attack size (which specifies the minimum initial attack that starts to make the system fail) and the critical attack size could vary by more than 50\% with different coupling coefficients. By adjusting the coupling coefficients, we could improve the robustness of the system.

However, such a fixed coupling coefficients strategy can still be improved. During the cascading failure process, if the coupling coefficients are fixed, and the initial conditions of the networks are determined (i.e., load distribution, free space distribution), and we also fix the redistribution strategy, then the fate of the system is determined by the initial attack.  However, the network conditions, including the fraction of the number of surviving nodes and the distribution of free space and load, 
when the cascading failure begins may be very different from its condition after a few steps of failures. For example, initially the network with more failed nodes may seem to be more vulnerable. After a few time steps, however, if we transfer more of this network's load to the other network, the other network may now be more vulnerable, i.e., it may have fewer surviving nodes and less free space 
compared to the one that experienced more initial failures. The optimal coupling coefficients at different times may thus be different. 
If network A is more vulnerable than the other ones, the optimal coupling coefficients might tend to redistribute A's extra load to other networks to generate less extra load in the following steps. After a few steps, if network A received less extra load in the previous steps, there might be some other network, say network B, that is more vulnerable compared to network A. Thus, we might want to adjust the coupling coefficients to add more extra load to network A instead of network B. Another example is:  Initially in a two network system, both networks have the same load distribution and network A has double the number of nodes compared to network B. To balance the flow/load in these two networks, the best coupling coefficients might result in around 67\% of extra load being redistributed in network A and 33\% of extra load being redistributed in network B. However, due to different free space distributions in the two networks, after a few steps, the number of surviving nodes may be similar in both networks. In this condition, the best coupling coefficients may result in the extra loads being equally redistributed in both networks from the viewpoint of system size. The robustness of the whole system could then be improved from the dynamic strategy to determine the coupling coefficients. 

Furthermore, for the fixed coupling coefficients strategy, there is no existing results suggesting a method to find the optimal settings of the coupling coefficients. A brute force search is used in \cite{zhang2018cascading} over all possible combinations of the coupling parameters, which may require much computation power. When the number of networks increases, determining the optimal coupling parameters might become too complicated since we may need to search for all possible combinations of coupling coefficients. 
Constructing a strategy to dynamically adjust the coupling coefficients may not only make the system more robust, but also allows us to more efficiently determine the coupling coefficients. We can set up a framework to determine the optimal coupling coefficients in a step-wise manner according to the recent system status without tracking the past status of each node in the system. Which means, during the cascading failure process, we could determine the optimal coupling coefficients to minimize the following extra load based on the system status and the average extra load it received in the previous time steps. 

\subsection{Two-Network Systems}

In this work, we will focus our discussion on two-network systems following \cite{zhang2018cascading, zhang2019robustness}, since it easily illustrates the idea of inter-dependencies of the networks. 
By dividing the extra load to be redistributed from each network to all $n$ networks with their corresponding coupling coefficients, we could simply extend the results to multiple networks. We will discuss this in the later section \ref{sec:discussion} 
In addition, for the dynamic strategy we proposed (discussed in detail in the next section), the difference between the 2 networks case and multiple networks case will be the number of coupling coefficients we need to solve. 

\begin{figure}[!h]
    \centering
    \includegraphics[width=1.0\linewidth]{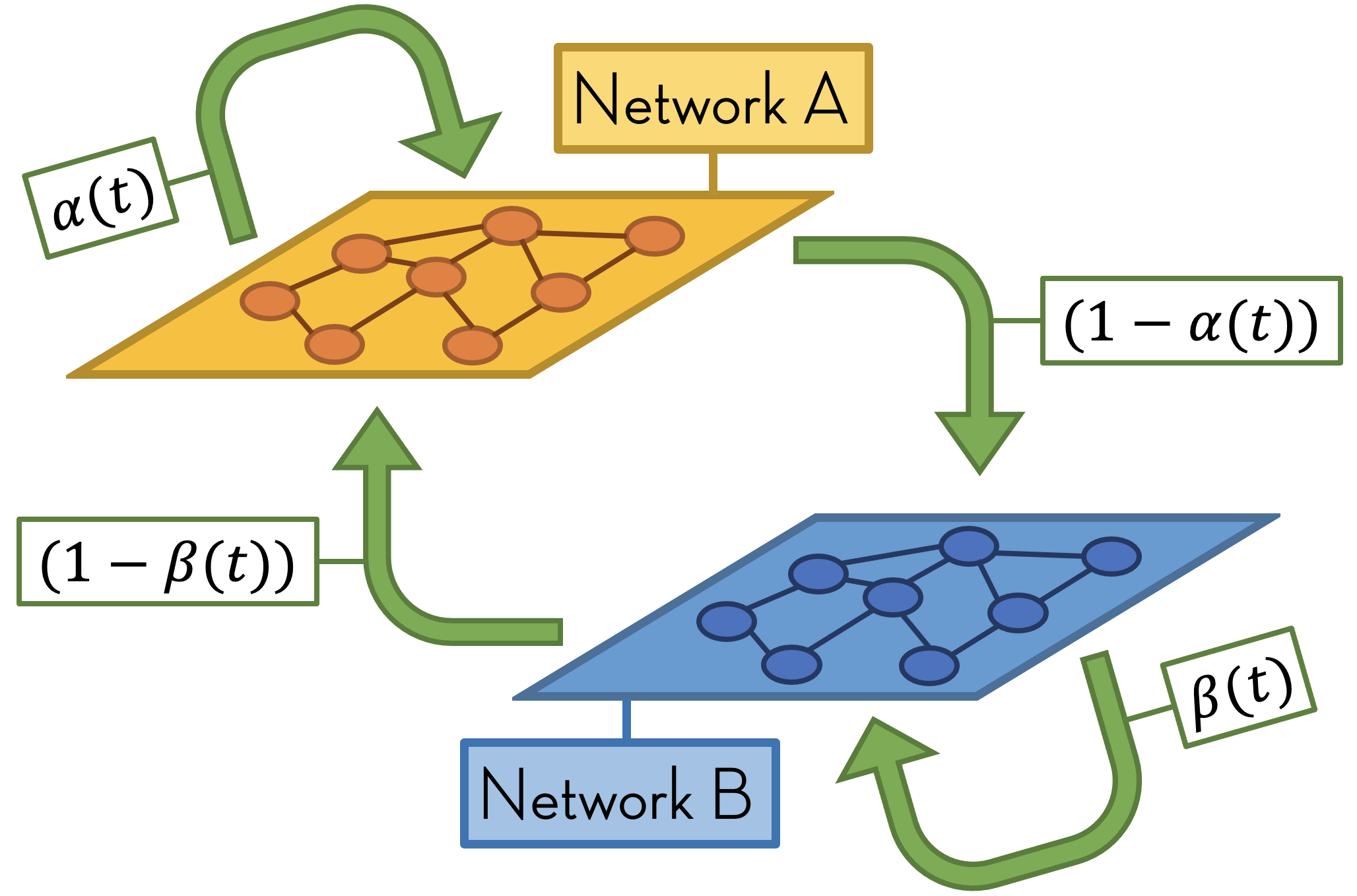}
    \caption{Illustration of the structure of the two networks system.}
    \label{fig:2NS}
\end{figure}

Consider a system with two networks, A and B. That is, $\mathcal{N} = \{ \mathcal{A}, \mathcal{B} \}$, initially each network has $N_A$ and $N_B$ number of nodes respectively and both networks are fully-connected. The size of the coupling matrix is now reduced to $2 \times 2$. For convenience, we define $\alpha (t)$ and $\beta (t)$ to be the portion of the extra load being redistributed internally for network A and network B at time $t$ respectively, which we also call the "in-net" portion for network A and network B. Thus, the portion of load being redistributed externally to another network at time $t$ will be $1-\alpha (t)$ and $1-\beta (t)$ respectively, we also called this as "out-net" portion, which are the coupling coefficients from network A to network B and from network B to network A. The initial failure probability (or fraction of nodes that initially fail) will be denoted as $p_A$ and $p_B$ for network A and network B respectively. Figure \ref{fig:2NS} illustrates the structure of this two-network system.

For a 2-network system, the parameters related to initial conditions are defined in Table \ref{IniParaDef}. We also define various time-dependent system parameters, which can be used to specify the status of the system. These parameters are listed in Table \ref{ParaDef} for network A. The corresponding parameters of network B have a ``B'' instead of ``A'' subscript. This notation will also be used for mathematical analysis and numerical evaluations in the following sections of this work. 

\begin{table}[hbt]
\begin{tabular}{ |p{1.2cm}|p{6.0cm}|  }
 \hline
 \multicolumn{2}{|c|}{List of parameters of the system at time $t$ for network A} \\
 \hline
 Notation & Definition \\
 \hline
 $t$ & Time step of the system. Initially $t=0$, after each iteration, the time proceed by 1 time step.\\
 \hline
$\alpha (t), \beta (t)$ & The in-net ratio at time $t$ for network A and B respectively.\\
 \hline
$N_{At}$ & Number of surviving nodes at time $t$ in network A. \\
 \hline
$f_{At}$ & Fraction of failed nodes up to time $t$ in network A. \\
 \hline
$F_{At}$ & Total extra load from network A at time $t$. \\
 \hline
$Q_{At}$ & Cumulative average extra load distributed at a single node in network A up to time $t$. \\
 \hline
$\Delta Q_{At}$ & Average extra load distributed at a single node in network A at time $t$. \\
 \hline
\end{tabular}
\captionof{table}{System Parameters at time $t$}\label{ParaDef}
\end{table}

\section{Mathematical Analysis}\label{sec:math}

Network systems like the transportation network in a greater metropolitan area of Tokyo or New York may consists of thousands or even millions of nodes. 
Tracking the dynamic variables of each node (i.e., its free space, load, and capacity) then might not be feasible in practical application and mathematical analysis. Thus, in this section, we provide a mean-field analysis of cascading failures in a two-network system, which reduces the number of required variables by characterizing the dynamics of the system in terms of average states and distributions across all nodes. 
With a large number of nodes, the mean-field analysis can specify the system status by tracking the surviving portion and average extra load of each network. The analysis was inspired by \cite{zhang2018cascading}, which we nontrivially extend to time-varying coupling coefficients. These present a novel challenge since the coupling coefficients change with each time step, altering the cascading failure process followed by the load redistribution at the most recent time step. 
This means we could not only use a simple constant portion to represent the interdependencies between networks. To iteratively derive the final status of the system, we cannot eliminate the terms in the middle of cascading failure process which was done in \cite{zhang2018cascading} with fixed coupling coefficients
, since the coupling coefficients now depend on time $t$ and are not the same for each time step. In addition, the strategy we proposed (which will be described in details later), require searching for optimal coupling coefficients at each time step and we need an efficient way to do so. 
Starting with the initiation of the cascading failure process, we provide a strategy for initializing and selecting the coupling coefficients at each time step. With the mathematical analysis in this section, we could calculate the final system size and dynamically optimize the coupling coefficients in a step-wise manner, solving the two critical challenges identified in the introduction.

\subsection{Initialization of Cascading Failure}

The cascading process starts from an initial attack, when a certain portion of the nodes in the network are attacked and fail. 
The load carried by these nodes is then spread to the other parts of the system, potentially causing other nodes to fail sequentially as described in Section 2.3. Following the notation listed in Table \ref{IniParaDef}, the following Definition \ref{def: initial} characterizes the initial condition of the network right after the initial attack happened.

\begin{definition}[Initial Conditions]
\label{def: initial}
Consider the system after the initial attack occurred in networks A and B, where the portions of nodes being attacked in each network are denoted by $p_A$ and $p_B$ respectively. For $\mathcal{X} \in \{A, B\}$, the status of the system, including the failing portion $f_{\mathcal{X}, 0}$, number of nodes $N_{\mathcal{X}, 0}$, total extra load $F_{\mathcal{X}, 0}$ and average extra load being redistributed per node $Q_{\mathcal{X}, 0}$ can be written as: 

\begin{equation}
    \begin{cases}
	f_{A0} & = p_A \\
	f_{B0} & = p_B \\
	N_{A0} & = (1-f_{A0})N_A \\
	N_{B0} & = (1-f_{B0})N_B \\
	F_{A0} & = N_A \cdot f_{A0} \cdot \mathbb{E}[L_A] \\
	F_{B0} & = N_A \cdot f_{B0} \cdot \mathbb{E}[L_B] \\
	Q_{A0} & = \Delta Q_{A0} = \frac{\alpha (0) \cdot F_{A0} + (1- \beta (0) \cdot F_{B0})}{(1-f_{A0}) \cdot N_A} \\
	Q_{B0} & = \Delta Q_{B0} = \frac{\beta (0) \cdot F_{B0} + (1- \alpha (0) \cdot F_{A0})}{(1-f_{B0}) \cdot N_B} \\
    \end{cases}
\end{equation}
Here $\alpha (0)$ and $\beta (0)$ denote the in-net redistribution portion 
at time $t=0$.

\end{definition}

The initial failed portion $f_{\mathcal{X} 0}$ is simply equal to the initial attack in network $\mathcal{X}$, since the cascading failure process has not started yet. Thus, the number of nodes surviving in  each network $\mathcal{X}$ simply equals the product of the initial number of nodes $N$ and the surviving portion, which equals $1-f_{\mathcal{X} 0}$. The total extra load to be redistributed from each network will be the sum of all loads of the failed nodes. Since the initial failure is fully random, in the mean-field sense, the total extra load to be redistributed could be derived by multiplying the expectation of the initial load and the expected number of failed nodes. The total extra load to be redistributed in a network will be the sum of loads from the same network and the load from the other network. Here $\alpha (0)$ and $\beta (0)$ determines the portion to be redistributed inside the same network and the portion to be redistributed to the other network. The average load to be redistributed per node, $Q_{\mathcal{X} 0}$, in both networks will be the total extra load to be redistributed in each network $\mathcal{X}$ divided by the number of surviving nodes in each network.

After the initial failure occurs, the cascading failure process is triggered if at least one of the two networks' average extra loads exceeds the free space of at least one node in the network. In such a case, at least one node will fail after the extra load from the initial failure is redistributed over the system.

If we denote the minimum free space among all nodes of network A and network B to be $S_{A0}$ and $S_{B0}$ respectively, utilizing the $Q_{A0}$ and $Q_{B0}$ from Definition \ref{def: initial} we can divide the initialization of the cascading failure process into three different cases:

\begin{itemize}
    \item \textbf{Case1}: $Q_{A0} < S_{A0}$ and $Q_{B0} < S_{B0}$. No cascading failure (CF) happens. The remaining part of the system stays intact.
    \item \textbf{Case2}: Either $Q_{A0} < S_{A0}$ or $Q_{B0} < S_{B0}$ but not both. Only one network starts the cascading failure process.
    \item \textbf{Case3}: $Q_{A0} \geq S_{A0}$ and $Q_{B0} \geq S_{B0}$. Both networks start the cascading failure process.
\end{itemize}

Once the cascading failure process starts, unless the internal redistribution ratio is set to 0, we could not stop the cascading failure process before it eventually reaches the equilibrium. In other words, case 1 will ends up without any cascading failure unless it received another attack. Once the cascading failure starts (case 2 or case 3), we could not stop the cascading failure before it reaches equilibrium, except in case 2, if the network where the cascading failure process is not triggered can absorb all the extra load being redistributed following the initial failure. 

Depending on the initial system parameters, we may be able to determine the initial values of $\alpha (0)$ and $\beta (0)$ to maximize the initial attack portion under which the system stays in case 1. In order to maximize the initial attack size. 
However, if the initial attack size is too large, the condition of case 1 can never be satisfied, and the optimal choice of $\alpha (0)$ and $\beta (0)$ will still remain unsolved, thus we will discuss this case in the following sections.

\subsection{Recursive Equations for Cascading Failure}

The status of each network will be based on the previous time step. Starting from step 1, the parameters specified in Definition \ref{def: initial} can be derived from Equations \ref{beginRC} to \ref{endRC} (explained below) and Theorem \ref{thm: recursive} recursively. For time $t=1$, the surviving portion will be the remaining portion after the initial attack times the probability that the free space of a node is greater than the average extra load after the initial attack. That is, the surviving portion $Sur_{A1}$ and $Sur_{B1}$ for network A and network B at time $t=1$ will be:

\begin{equation}
\label{beginRC}
\begin{cases}
Sur_{A1} = (1-p_A)\cdot P[S_A \geq Q_{A0}] \\
Sur_{B1} = (1-p_B)\cdot P[S_B \geq Q_{B0}] \\
\end{cases}
\end{equation}

Thus, the failed fractions of nodes $f_{A1}$ and $f_{B1}$ for network A and network B are simply one minus the surviving portion, that is:

\begin{equation}
\begin{cases}
f_{A1} = 1- Sur_{A1} = 1- (1-p_A)\cdot P[S_A \geq Q_{A0}] \\
f_{B1} = 1- Sur_{B1} = 1- (1-p_B)\cdot P[S_B \geq Q_{B0}]. \\
\end{cases}
\end{equation}

The number of surviving nodes at time step 1 is then simply the surviving portion times the number of initial nodes, that is:

\begin{equation}
\begin{cases}
N_{A1} = Sur_{A1} \cdot N_A = (1-p_A)\cdot N_A \cdot P[S_A \geq Q_{A0}] \\
N_{B1} = Sur_{B1} \cdot N_B = (1-p_B)\cdot N_B \cdot P[S_B \geq Q_{B0}]. \\
\end{cases}
\end{equation}

The expected total extra load to be redistributed at $t=1$ for both networks is the product of the number of failing nodes at the current time step and the expected load carried by a single node. That is:

\begin{equation}
\begin{cases}
F_{A1} & = f_{A1} \cdot N_A \cdot \mathbb{E}[L_A+Q_{A0}] \\
& = (1-(1-p_A)\cdot P[S_A \geq Q_{A0}])\cdot N_A \cdot P[S_A \geq Q_{A0}] \\
F_{B1} & = f_{B1} \cdot N_B \cdot \mathbb{E}[L_B+Q_{B0}] \\ 
& = (1 - (1-p_B)\cdot P[S_B \geq Q_{B0}])\cdot N_B \cdot P[S_B \geq Q_{B0}] \\
\end{cases}
\end{equation}

Finally, the average extra load for network A and network B at $t=1$ will be the sum of internal extra load and extra load from the other network, divided by the number of surviving nodes in the network. We already have the total extra loads and the number of surviving nodes, thus we can write the average extra load per node at time $t=1$ as:

\begin{equation}
\begin{cases}
\Delta Q_{A1} = \frac{\alpha (1) \cdot F_{A1} + (1-\beta(1))\cdot F_{B1}}{N_{A1}} \\
\Delta Q_{B1} = \frac{\beta (1) \cdot F_{B1} + (1-\alpha(1))\cdot F_{A1}}{N_{B1}}\\
\end{cases}
\end{equation}

The cumulative average extra load per node can be derived by simply adding the average extra load at the current time $t=1$ and the average extra load per node at previous time step $t=0$:

\begin{equation}
\label{endRC}
\begin{cases}
Q_{A1} = \Delta Q_{A1} + Q_{A0} \\
Q_{B1} = \Delta Q_{B1} + Q_{B0} \\
\end{cases}
\end{equation}

Following the same manner, starting from time step $t \geq 2$, we have the following Theorem \ref{thm: recursive} specifying the recursive equations for the system.

\begin{theorem}[Recursive Equations for the System]
\label{thm: recursive}
For time step $t \geq 2$, the status of the system can be written as the following recursive equations.

For network A, we have:
\begin{equation}
    \begin{cases}
	f_{At} & = 1 - (1-p_A) \cdot P[S_A \geq Q_{A(t-1)}] \\
	N_{At} & = (1-f_{At})N_A \\
	F_{At} & = N_A \cdot (f_{At}-f_{A(t-1)}) \cdot \mathbb{E}[L_A + Q_{A(t-1)}] \\
	& = N_A \cdot (1-p_A) \\
	& \cdot P[Q_{A(t-2)} < S_A \leq Q_{A(t-1)}] \cdot \mathbb{E}[L_A + Q_{A(t-1)}] \\
	\Delta Q_{At} & = \frac{\alpha (t) \cdot F_{At} + (1- \beta (t) \cdot F_{Bt})}{(1-f_{At}) \cdot N_A} \\
	Q_{At} & = Q_{A(t-1)} + \Delta Q_{At} \\
    \end{cases}
\end{equation}

For network B, we have:
\begin{equation}
    \begin{cases}
	f_{Bt} & = 1 - (1-p_B) \cdot P[S_B \geq Q_{B(t-1)}] \\
	N_{Bt} & = (1-f_{Bt})N_B \\
	F_{Bt} & = N_B \cdot (f_{Bt}-f_{B(t-1)}) \cdot \mathbb{E}[L_B + Q_{B(t-1)}] \\
	& = N_B \cdot (1-p_B) \\
	& \cdot P[Q_{B(t-2)} < S_B \leq Q_{B(t-1)}] \cdot \mathbb{E}[L_B + Q_{B(t-1)}] \\
	\Delta Q_{Bt} & = \frac{\beta (t) \cdot F_{Bt} + (1- \alpha (t) \cdot F_{At})}{(1-f_{Bt}) \cdot N_B} \\
	Q_{Bt} & = Q_{B(t-1)} + \Delta Q_{Bt} \\
    \end{cases}
\end{equation}

The status of the system at time $t$ can thus be derived iteratively following the recursive equations, starting from time step $t=0$.

\end{theorem}

The proof of Theorem \ref{thm: recursive} follows the same outline as the derivation of the network status at time step $t=1$ in Equations (\ref{beginRC}) to (\ref{endRC}). The failing portion at time $t$ can be directly found from the total size of the network, less the surviving portion of the network. The number of surviving nodes is simply the product of surviving portion and the initial number of nodes. The total extra load is derived from the expected load a single node at $t-1$ 
have multiplied by the expected number of nodes being failed at previous time step. Then the (cumulative) average extra load is simply the total load received by the network, divided by the number of surviving nodes in the network.

From the above recursive equations, we can calculate the parameters step by step and eventually get the final state of the system after the cascading failure process. The cascading failure process only stops when the system size (i.e. number of nodes) does not change, i.e., when:

\begin{equation}
    \begin{cases}
	N_{At} - N_{A(t-1)} \longrightarrow 0\\
	N_{Bt} - N_{B(t-1)} \longrightarrow 0 \\
    \end{cases}
\end{equation}

As stated in Section 2.3, there are two stopping conditions for the cascading failure process. Either the whole system breaks down (i.e., no node in the system survives), or an equilibrium subset of nodes will survive.

\subsection{Dynamic Coupling Coefficient Strategies: Step-wise Optimization (SWO)}
The recursive equations for the network conditions in Theorem \ref{thm: recursive} allow us to evaluate the system numerically. However, our goal is to find the optimal coupling coefficients in each time step. The choice of the optimal coupling coefficients ($1-\alpha(t)$ and $1-\beta(t)$), and thus the ratio being redistributed inside the same network (this is the in-net ratio ($\alpha(t)$ and $\beta(t)$), at each time step requires further analysis. We now provide the intuitive reasoning by which we determine the optimal values of $\alpha (t)$ and $\beta (t)$ at each time step $t$, which we formalize in Definition~\ref{def:opt}. 
The choices of $\alpha(t)$ and $\beta(t)$ for the current time step $t$ directly influence the portion of the extra load added to both networks. This extra load will determine the average extra load redistributed to each node in both networks at the next time step. 
The amounts of extra load at the current time step redistributed to other nodes will determine the number of nodes, and more specifically which nodes, will fail in both networks during the next time step. In other words, the coupling coefficients determine how much extra load must be absorbed by each network. According to the current system status (Theorem~\ref{thm: recursive}), this determines which nodes are failed, and then the total extra load to be redistributed at the next time step is also determined. The direct consequence of the choice of coupling coefficients is then the number of nodes that fail in the next time step and the total extra load to be redistributed at the next time step. 

If the system ultimately breaks down, then the cascading failure will stop when all nodes have failed. Equivalently, all of the nodes fail when the cumulative extra load from each time step exceeds nodes' total available free space. 
The "limited" resource of the system are the number of nodes and the total free space. To avoid the whole system breaking down, we would thus want the amount of extra load and the number of failed nodes in each time step to be as small as possible. 

The above intuition suggests that our choice of coupling coefficients should minimize either the number of nodes that will fail at the next time step, or the total extra load to be redistributed. Either of these will make stopping the cascading failure more likely. If we only consider the number of nodes that fail, however, then we ignore differences in the load that every node carries and the free space that every node can provide. Failing a node with more free space or less load is different from failing a node with less free space or more load. The first case indicates that the system loses more capability to accommodate future extra load and poses less immediate pressure to the system compared to failing nodes with less free space. Simply considering the number of surviving nodes does not discriminate between these scenarios. 
A strategy that minimizes the number of nodes failing at the next time step 
only works well when we consider the special case that both networks in the system are identical, that is, having the same initial load and free space distribution for all networks. In later section 3.4 the introduced method size-based dynamic is one of the special case of minimizing number of failing node under uniform free space distribution which we use to compare with our proposed method. 
However, for a more general case, the load/free space viewpoint will likely perform better. 

To take the load and free space into consideration, our step-wise strategy will aim to minimize the total extra load to be redistributed at the next time step. From our recursive equations in Theorem \ref{thm: recursive}, in the mean field analysis the total extra load at the next time step can be written as the product of a node's expected load and the number of nodes that will fail at the next time step. We thus aim to minimize the sum of this quantity for both networks in the system. The constraints of the optimization problem are that the coupling coefficients should be feasible. For systems without any further constraints, the coupling coefficient should be between 0 and 1.  
Thus, our optimization problem can be written in the following Definition \ref{def: opt}:

\begin{definition}[The Step-wise Optimization Problem]
\label{def: opt}
In each step of the cascading failure process, greedily minimize the expected extra load in the next time step by adjusting the coupling coefficients at the current time step. We can formulate the optimization problem at each time step as:
\begin{equation}
   \min _{\alpha, \beta}\left(\sum_{i=A, B}\mathbb{E}\left[L_{i}+Q_{i(t+1)}\right] \Delta N_{i(t+1)}\right) \\ 
\end{equation}

subject to:

\begin{equation}
    \begin{cases}
        0 \leq \alpha (t) \leq 1 \\
        0 \leq \beta (t) \leq 1 \\
    \end{cases}
\end{equation}
\end{definition}

Some real systems may have more constraints, for example, in a transportation system, a coupling coefficient of 1 from the subway network to the bus network means that we need to get all passengers to switch from subway to buses, which is unrealistic. Depending on the design of different system, the upper-bound of the coupling coefficients may be some value less than 1. We can simply modify the constraints in Definition \ref{def: opt} to adapt to these kinds of system.

All the quantities in Definition \ref{def: opt} only depend on the initial settings of the networks and the status of the most recent time steps of the network.

The expectation of the load on a single node can be written as:

\begin{equation}
   \begin{cases}
    \mathbb{E}\left[L_{A}+Q_{A(t+1)}\right] &=\mathbb{E}\left[L_{A}+Q_{A t}+\Delta Q_{A(t+1)}\right] \\
    \mathbb{E}\left[L_{B}+Q_{B(t+1)}\right.&=\mathbb{E}\left[L_{B}+Q_{B t}+\Delta Q_{B(t+1)}\right] \\
    \end{cases} 
\end{equation}

Here $L_A$ and $L_B$ depend on the initial settings of the system, and $Q_{At}$ and $Q_{Bt}$ are the cumulative average extra load on a single node up to current time step. $\Delta Q_{A(t+1)}$ and $\Delta Q_{B(t+1)}$ depends on the coupling coefficients $\alpha (t)$ and $\beta (t)$ which can be written as: 

\begin{equation}
    \begin{cases}
        \Delta Q_{A(t+1)}=\frac{\left(\alpha(t) Q_{A t} N_{A t}+(1-\beta(t)) Q_{B t} N_{B t}\right)}{N_{A t}} \\
        \Delta Q_{B(t+1)}=\frac{\left(\beta(t) Q_{B t} N_{B t}+(1-\alpha(t)) Q_{A t} N_{A t}\right)}{N_{B t}} \\
    \end{cases}
\end{equation}

On the other hand, $ \Delta N_{A(t+1)}$ and $ \Delta N_{B(t+1)}$ are the number of nodes failing at the next time step, which can be written as:

\begin{equation}
    \begin{cases}
        \Delta N_{A(t+1)} = (1-p_A) \cdot N_A \cdot P[Q_{A(t-1)} \leq S_A \leq Q_{At}] \\
        \Delta N_{B(t+1)} = (1-p_B) \cdot N_B \cdot P[Q_{B(t-1)} \leq S_B \leq Q_{Bt}] \\
    \end{cases}
\end{equation}

Analytically, these equations depend on the free space distribution $S_A$ and $S_B$ of the networks. Practically, if we know 
the distribution of the free space, we can also use it to estimate the number of nodes failing at next time step.

\subsection{Special Case: Identical Uniform Distribution}

In order to illustrate how our step-wise optimization strategy works, we use the special case of identical networks with uniform free space distribution. This case gives a simple closed form of the global optimal point at each time step, since the number of surviving nodes could be directly related to the average extra load which we used in the mean-field analysis. We set the initial load on every node in both network A and network B to be $L$, and the free space of both networks follow the same uniform distribution, which is:

\begin{equation}
    \begin{cases}
        S_A \sim U (S_0, S_0 +d)\\
        S_B \sim U (S_0, S_0 +d)\\
    \end{cases}
\end{equation}

As stated above, the mathematical property of uniform distribution simplifies the objective function of the step-wise optimization problem. This is because the fraction of nodes that survive is directly related to fraction of the range between the minimum free space distribution, the cumulative average extra load $Q_{\bullet t}-S_0$, and the range $d$ of the uniform distribution. Following the optimization problem defined in the previous section in Definition \ref{def: opt}, we have the optimization problem for this special case as the following Definition \ref{def: optid}:

\begin{definition}[The Step-wise Optimization Problem for Identical Networks with Uniform Distribution]
\label{def: optid}
The optimization problem at each time step with two identical networks with uniform distribution can be written as:
\begin{equation*}
    \begin{array}{cc}
         \min _{\alpha(t), \beta(t)} & (N_{A}\left(1-p_{A}\right) \cdot \delta_{A}\left[\delta_{A}+\left(L-Q_{A(t-1)}\right)\right]+ 
         \\& N_{B}\left(1-p_{B}\right) \cdot \delta_{B}\left[\delta_{B}+\left(L-Q_{B(t-1)}\right)\right]) \\
    \end{array}
\end{equation*}

Subject to:

\begin{equation}
    \begin{cases}
        0 \leq \alpha (t) \leq 1 \\
        0 \leq \beta (t) \leq 1 \\
    \end{cases}
\end{equation}

Where:

\begin{equation}
    \begin{cases}
        \delta_{A} = & \frac{\alpha(t)\left(L+Q_{A(t-1)}\right) \Delta Q_{A(t-1)}}{d\left(S_{0}+d-Q_{A t}\right)} \\
        & + \frac{(1-\beta(t))\left(L+Q_{B(t-1)}\right) \Delta Q_{B(t-1)}\left(1-p_{B}\right) N_{B}}{\left(1-p_{A}\right) d N_{A}\left(S_{0}+d-Q_{A t}\right)} \\
        \delta_{B} = & \frac{\beta(t)\left(L+Q_{B(t-1)}\right) \Delta Q_{B(t-1)}}{d\left(S_{0}+d-Q_{B t}\right)} \\
        & + \frac{(1-\alpha(t))\left(L+Q_{A(t-1)}\right) \Delta Q_{A(t-1)}\left(1-p_{A}\right) N_{A}}{d\left(S_{0}+d-Q_{B t}\right)\left(1-p_{B}\right) N_{B}} \\
    \end{cases}
\end{equation}
\\
\end{definition}

Since the objective function is quadratic, the global optimal $\alpha ^*(t)$ and $\beta ^*(t)$ can be written in a closed form which is shown in Theorem \ref{thm: optid}.

\begin{theorem}
[The Solution of Optimal Coupling Coefficients for Identical Networks with Uniform Distribution]
\label{thm: optid}
The global optimal $\alpha ^*(t)$ and $\beta ^*(t)$ can be written as:

\begin{equation}
    \alpha^{*}(t)=\frac{\frac{K_{\alpha \beta}^{2}}{2 K_{\beta 2}}-2 K_{\alpha 2}}{K_{\alpha}-\frac{K_{\alpha \beta} K_{\beta}}{2 K_{\beta 2}}}, \quad \beta^{*}(t)=\frac{\frac{K_{\alpha \beta}^{2}}{2 K_{\alpha 2}}-2 K_{\beta 2}}{K_{\beta}-\frac{K_{\alpha \beta} K_{\alpha}}{2 K_{\alpha 2}}} \\
\end{equation}

Where:

$$
\begin{array}{l}
K_{\alpha 2}=A_{1}^{2} N_{A}^{\prime}+B_{2}^{2} N_{B} \\
K_{\beta 2}=B_{1}^{2} N_{A}^{\prime}+A_{2}^{2} N_{B} \\
K_{\alpha}=A_{1} N_{A}^{\prime}\left(B_{1}+L_{A}\right)-B_{2} N_{B}\left(B_{2}+L_{B}\right) \\
K_{\beta}=-B_{1} N_{A}^{\prime}\left(B_{1}+L_{A}\right)+A_{2} N_{B}\left(B_{2}+L_{B}\right) \\
K_{\alpha \beta}=-2\left(N_{A}^{\prime} A_{1} B_{1}+N_{B} A_{2} B_{2}\right) \\
\end{array}
$$

and
\begin{equation}
    \begin{cases}
    \mathrm{A}_{1}=\frac{\left(L_{A}+Q_{A(t-1)}\right) \Delta Q_{A(t-1)}}{d\left(S_{0}+d_{A}-Q_{A t}\right)}\\ 
    B_{1}=\frac{\left(L_{B}+Q_{B(t-1)}\right) \Delta Q_{B(t-1)}\left(1-p_{B}\right) N_{B}}{\left(1-p_{A}\right) d N_{A}\left(S_{0}+d-Q_{A t}\right)} \\
    \mathrm{A}_{2}=\frac{\left(L_{B}+Q_{B(t-1)}\right) \Delta Q_{B(t-1)}}{d\left(S_{0}+d-Q_{B t}\right)} \\
    B_{2}=\frac{\left(1-p_{A}\right)\left(L_{A}+Q_{A(t-1)}\right) \Delta Q_{A(t-1)} N_{A}}{d\left(S_{0}+d-Q_{B t}\right)\left(1-p_{B}\right) N_{B}} \\
    \end{cases}
\end{equation}
\\
\end{theorem}

To show that the optimization problem in Definition \ref{def: optid} is convex and the solution of optimal coupling coefficients in Theorem \ref{thm: optid} is valid, we focus on the quadratic terms of the objective function. Where the objective function could be written as:

\begin{equation}
\begin{matrix}
    obj = & - 2\alpha(t)\beta(t) (N_{A0}A_1 B_1 + N_{B0}A_2 B_2) \\
    & + \alpha(t)^2 (N_{A0}A_1^2+N_{B0}B_2^2) \quad \quad \quad \quad \quad\\
    & + \beta(t)^2 (N_{A0}B_1^2+N_{B0}A_2^2) \quad \quad \quad \quad \quad\\
    & + ... \quad \quad \quad \quad \quad \quad \quad \quad \quad \quad \quad \quad \quad \quad \quad\\
    \end{matrix}
\end{equation}

The Hessian of the objective function will be:

\begin{equation}
\begin{array}{l}
\nabla ^2 obj = \\
\\
\begin{pmatrix}
    2(N_{A0}A_1^2+N_{B0}B_2^2) & -2(N_{A0}A_1 B_1 + N_{B0}A_2 B_2)\\
    -2(N_{A0}A_1 B_1 + N_{B0}A_2 B_2) &  2(N_{A0}B_1^2+N_{B0}A_2^2) \\
\end{pmatrix} \\
\end{array}
\end{equation}

In the special case of 2 by 2 matrix, 
this implies that the matrix is positive semi-definite, since the trace equal to the sum of eigenvalues is positive, the determinant equal to the product of eigenvalues is positive. Thus, the eigenvalues of the Hessian matrix are positive, and the objective function is convex.

For such quadratic objective function with uniform distribution, we have the closed form for the global minimum, so we can simply apply the equation in Theorem \ref{thm: optid} to find the solution. If the global minimum does not lie inside the set of feasible solutions defined in Definition \ref{def: optid}, the minimum point inside the feasible set should lie at the boundary of the set. We can then check four different cases where $\alpha (t) = 0$ or $1$ or $\beta (t) = 0$ or $1$, the objective function will become a single-variable quadratic function. We then check if the minimum point lies inside the range $0 \leq \beta (t) \leq 1$ for the first two cases and $0 \leq \alpha (t) \leq 1$ for the last two cases. At last, compare these values with the four vertices of the feasible set $(\alpha (t), \beta (t)) = (0, 0), (0, 1), (1, 0), (1, 1)$ to find the minimum value of the objective function and determine the optimal coefficients $\alpha (t)$ and $\beta (t)$.

To evaluate our SWO strategy, we compare it to a heuristic strategy that we call the size-based dynamic (SBD) strategy. We construct a model with a random key graph \cite{yagan2012zero,RKG2009} where each node has different keys, and nodes with same key are assumed to be connected. Thus we could view the nodes with the same keys as being in the same fully-connected network. We have shown the analytical results of random key graphs under identical settings in the appendices, 
under the identical settings, when the number of nodes is large enough, setting the time-varying coupling coefficients directly equal to the ratio of surviving nodes will let each node in the system receive same extra load in each time step. This scenario thus resembles a single fully-connected 
network that all nodes are viewed equally and being connected to all other nodes under a single network, all the load redistribution process is the same as a equally redistribution over a single fully-connected network. In other words, in this case that all nodes in the system received the same extra load at each time step, which intuitively could be viewed as a single fully-connected network. This is a special case of minimizing number of nodes being failed at the next step under uniform free space distribution in our settings as we mentioned in section 3.3. 

We will compare the results from the step-wise optimization (SWO) strategy we proposed and the results from this single fully-connected like strategy, which we called this Size-based dynamic (SBD) coupling strategy in the following discussion. It can guarantee the performance of our SWO strategy if it can perform equally good or better than the SBD strategy which is known to exhibit a great robustness against cascading failure.

We finally note that any free space distribution for which the objective function in Definition \ref{def: opt} is convex can easily employ this framework to solve for the optimal values of $\alpha ^*(t)$ and $\beta ^*(t)$ at each time step $t$. For example, the exponential distribution results in a convex objective.  This is simply due to the memoryless property of the Exponential distribution: if the initial load distribution is Exponential, the load distribution of the surviving nodes at each time step is also an Exponential distribution with origin shifting to the current cumulative average extra load $Q_{At}$ and $Q_{Bt}$. In case the objective function is not convex, under a two-network system, we only have 2 optimization variables, we could still apply grid search for different combinations of coupling coefficients.

\section{Numerical Results}\label{sec:num}

To verify the performance of our proposed step-wise optimization (SWO) method, we evaluate its performance compared to baseline algorithms on different kinds of free space distributions under different network settings. We show that SWO is as good as or better than SBD and fixed coupling in a variety of settings. 

To evaluate the performance of our SWO method, the structure of this section will be as follows: We begin with the special case of two identical networks, i.e., both networks in the system have the same number of nodes and load/free space distributions. In Section 4.1, we compare the performance of our SWO method to that of the SBD method when the networks are identical, showing that our SWO method could perform as well as SBD. In Section 4.2, we then compare the SBD method to strategies with fixed coupling coefficients, again under identical network settings, which shows that our methods outperform the fixed coupling coefficients (FCC) strategies. 
To verify the correctness of our analysis, 
we compare the simulation results with the results derived from the analytical equations from Definition \ref{def: initial} and Theorem \ref{thm: recursive} in Section 4.3, which validates our mean-field analysis. Then, in Section 4.4 we show that our SWO method outperforms the SBD method when the networks are not identical. When the networks are not identical, the SBD method may not work as well since this strategy views all nodes in each network as the same. 
Our SWO strategy, however, takes the load and free space into consideration. On the other hand, our SWO method can outperform the optimal FCC settings. Finally, we extend our SWO method to networks that are not fully connected 
in Section 4.5, demonstrating that our SWO method continues to perform well 
when the structure of the network is not fully-connected and the load redistribution will follow the network topology. The results shows that our SWO method outperforms most settings of fixed-coupling coefficients, but could be worse than some settings due to incorrect estimation of the nodes' free space.

\subsection{Identical Networks: SWO and SBD}
We first compare our SWO method with the SBD method to show that our method could perform better than or at least as well as the SBD method.

Under the identical network setting, both networks have exactly the same load and free space distributions. Inspired by the results from the random key graph (RKG) (see appendix), we introduce the SBD strategy that mimics the behavior of the equally redistribution strategy under a single fully-connected network that we explained in Section 3.4. 
This is expected to perform well since all nodes in the system share the load together, and equal redistribution means that it does not cause any part of the network tend to fail earlier than the other parts. 
This strategy will ensure that the two networks have the same free space and load distributions at each time step. It only depends on the number of surviving nodes and sets the coupling coefficients as follows:

\begin{equation}
\begin{cases}
    \alpha (t) = \frac{n_t(A)}{n_t(A)+n_t(B)} \\ 
    \beta (t) = \frac{n_t(B)}{n_t(A)+n_t(B)} \\ 
\end{cases}
\end{equation}

Here $n_t(A)$ and $n_t(B)$ represent the number of surviving nodes of network A and network B at time step $t$.

By using this strategy, the original recursive equations in Theorem \ref{thm: recursive} 
of average extra load for time step $t$ can be written as:

\begin{equation}
\begin{cases}
	\Delta Q_{At} = \Delta Q_{Bt} = \frac{F_{At} + F_{Bt})}{N_{At} + N_{Bt}} \\
	Q_{At} = Q_{A(t-1)} + \Delta Q_{At} \\
	Q_{Bt} = Q_{B(t-1)} + \Delta Q_{Bt} \\
\end{cases}
\end{equation}

This SBD strategy will redistribute the extra load equally to the nodes in each time step. The identical networks system using SBD strategy is statistically identical to the single network case 
with attack size being the weighted average of the initial attack size of two networks. It has exactly the same behavior as equal redistribution in a single fully-connected network.

Unless otherwise stated, in all of our experiments the system contains two networks. Each network has the size of 1 million nodes, $N_A = N_B = 10^6$. For each data point in our figures (i.e., each setting at different initial attack size), the value is taken from the average of 100 independent experiments.

\begin{figure}[!h]
    \centering
    \includegraphics[width=1.0\linewidth]{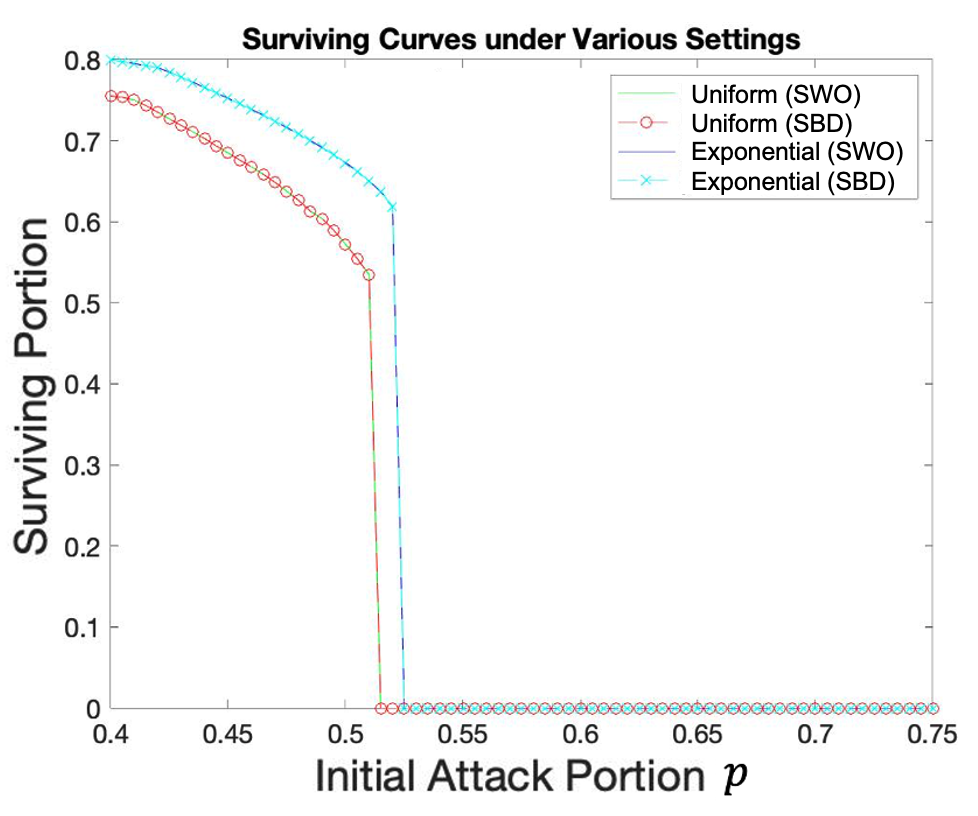}
    \caption{Simulation of different strategies for a system with identical networks. Our step-wise optimization strategy and the size-based dynamic strategy have almost the same performance.}
    \label{fig:IDN}
\end{figure}

The results for identical networks are shown in Figure \ref{fig:IDN}. The four different curves use either SWO or SBD strategy with uniform distribution or exponential distribution of free space, which are: 

\begin{itemize}
    \item \textbf{Uniform: } $S_A, S_B \sim U(20, 180)$, $\mathbb{E}[L_A] = \mathbb{E}[L_B] = 75$. 
    \item \textbf{Exponential: } $S_A, S_B \sim 20 + Exponential(\frac{1}{120})$, $\mathbb{E}[L_A] = \mathbb{E}[L_B] = 60$.
    \\
\end{itemize}

Under the identical network case, our SWO strategy has the same performance as the SBD strategy under different settings, in terms of remaining portion of the system and the critical attack size. In other words, our SWO works as well as the SBD strategy.

\begin{figure}[!h]
    \centering
    \includegraphics[width=1.0\linewidth]{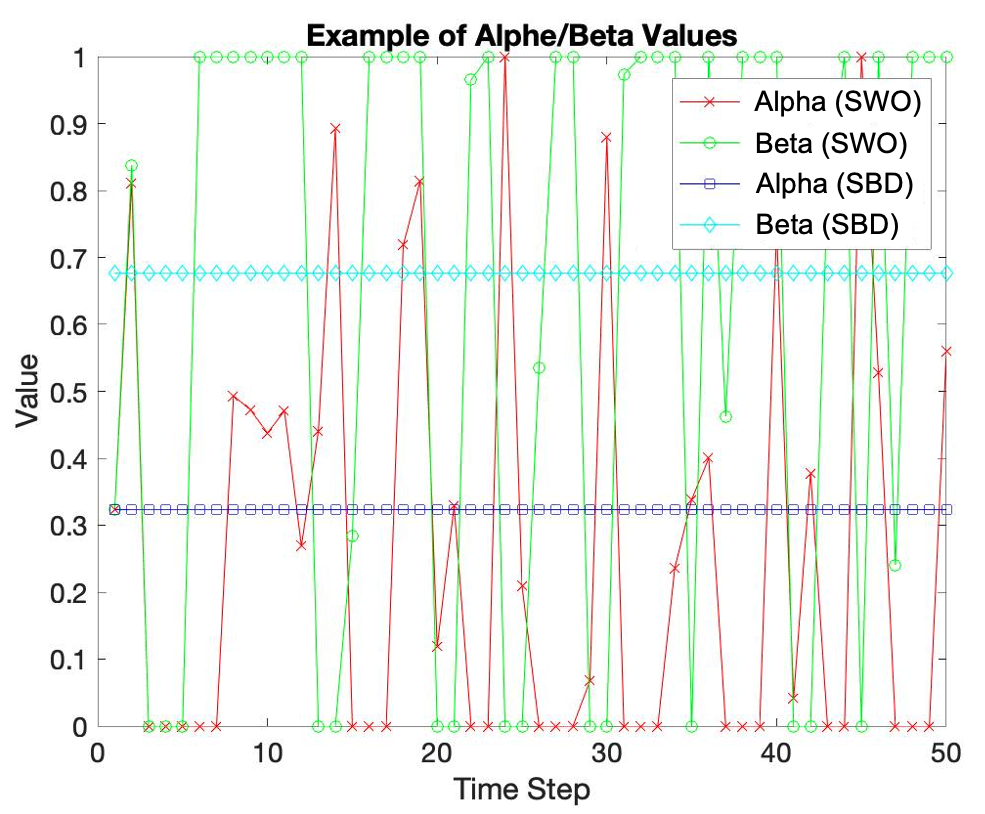}
    \caption{Example of values of $\alpha$ and $\beta$ for identical networks.}
    \label{fig:AB}
\end{figure}

However, if we look into the values of $\alpha (t)$ and $\beta(t)$ that our SWO strategy takes in each time step $t$, we see a great difference with the SBD strategy. Following the discussion in Section 3.4, we know that if the global minimum point does not lie in the feasible region, the coupling coefficients will take values on the boundary. Figure \ref{fig:AB} shows an example of the values of $\alpha$ and $\beta$ during each time step of the cascading failure process using uniform distribution of the free space. The SWO strategy often takes coupling coefficient values at the boundary of $\alpha,\beta = 0,1$ and sometimes takes values in between. The global minimum of the objective function in each time step is usually outside the feasible region. 
The values of coupling coefficients for the SBD strategy take almost constant values around 0.5 (in fact, it tries to maintain the ratio of the number of the nodes in network A and B, taking value around the ratio of the number of nodes in network A and B after initial failure).

Our SWO strategy tends to find the minimum points at each step, it tend to fix the error between the expected extra load and the real extra load being redistributed determined in the previous time step 
at each step according to current network situations.

\subsection{Identical Networks: SWO/SBD and Fixed Coupling Coefficients (FCC) Strategy}

We have shown that our SWO strategy has the same performance as the SBD strategy under identical networks settings. We now compare these to the fixed coupling coefficients (FCC) strategy, where the coupling coefficients are the same through out the whole cascading failure process. 

As in the previous simulation, each network has the size of 1 million nodes, $N_A = N_B = 10^6$. For each data point (i.e. each settings at different initial attack size), the value is taken from the average of 100 independent experiments.

\begin{figure}[!h]
    \centering
    \includegraphics[width=1.0\linewidth]{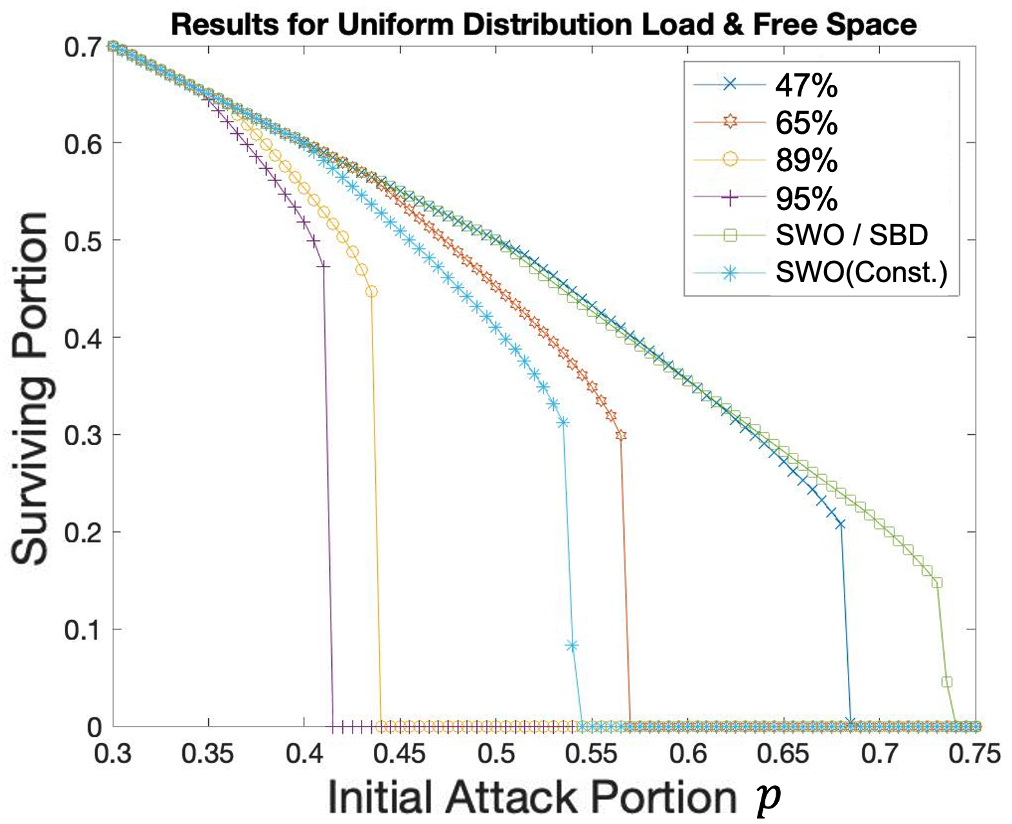}
    \caption{Simulation for different strategies for system with identical networks. The SWO/SBD strategies is shown to outperform the FCC strategies.}
    \label{fig:CRF}
\end{figure}

We now follow the settings where $L_A, L_B \sim U(10, 30)$ and $S_A, S_B \sim U(10, 65)$. The results for the SWO/SBD strategy and the FCC strategy (where $\alpha (t) = \beta (t)$) are shown in the Figure \ref{fig:CRF}.
Our SWO method not only yields good performance, but can also easily accommodate constraints on the coupling coefficients.
If we add a constraint to the coupling coefficients by setting the $\alpha, \beta \in [0.5, 1]$, for example, which constrains the fraction of load going to the other network, we show in Figure \ref{fig:CRF} that we can still have a good performance on our system. 
The meanings of the labels in the figure are:

\begin{itemize}
    \item \textbf{FCC ($X \%$): } The FCC strategies where the $X \%$ indicates the portion of extra load to be redistributed in the same network, that is, $\alpha (t) = \beta (t) = X \%$. Where the $65\%$ case is the best among all other possible choices of fixed strategies.
    \item \textbf{SBD/SWO: } The SBD strategy and SWO strategy (which have the same curve).
    \item \textbf{SWO (Const.): } The SWO strategy with constraints $\alpha (t), \beta (t) \in [0.5, 1]$. 
    \\
\end{itemize}

The experiment results in Figure \ref{fig:CRF} show that the {SBD}/SWO strategy has the best robustness among all other FCC strategies. Thus under identical network settings, our SWO strategy outperforms all FCC strategies. For the constrained SWO strategy, it could still outperform some of the fixed strategy even if we posed constraints to the coupling coefficients. 

\subsection{Verification: Analytical Equations vs. Simulation}

To verify the correctness of the previous simulations, we compare the simulation results with the numerical results from the analytical equations (Definition \ref{def: initial} and Theorem \ref{thm: recursive} in Section 4.3). 
If the two results are the same, we can make sure that taking the average among all nodes in the mean-field analysis does not affect the accuracy of our results compared to the real situation. 

\begin{figure}[!h]
    \centering
    \includegraphics[width=1.0\linewidth]{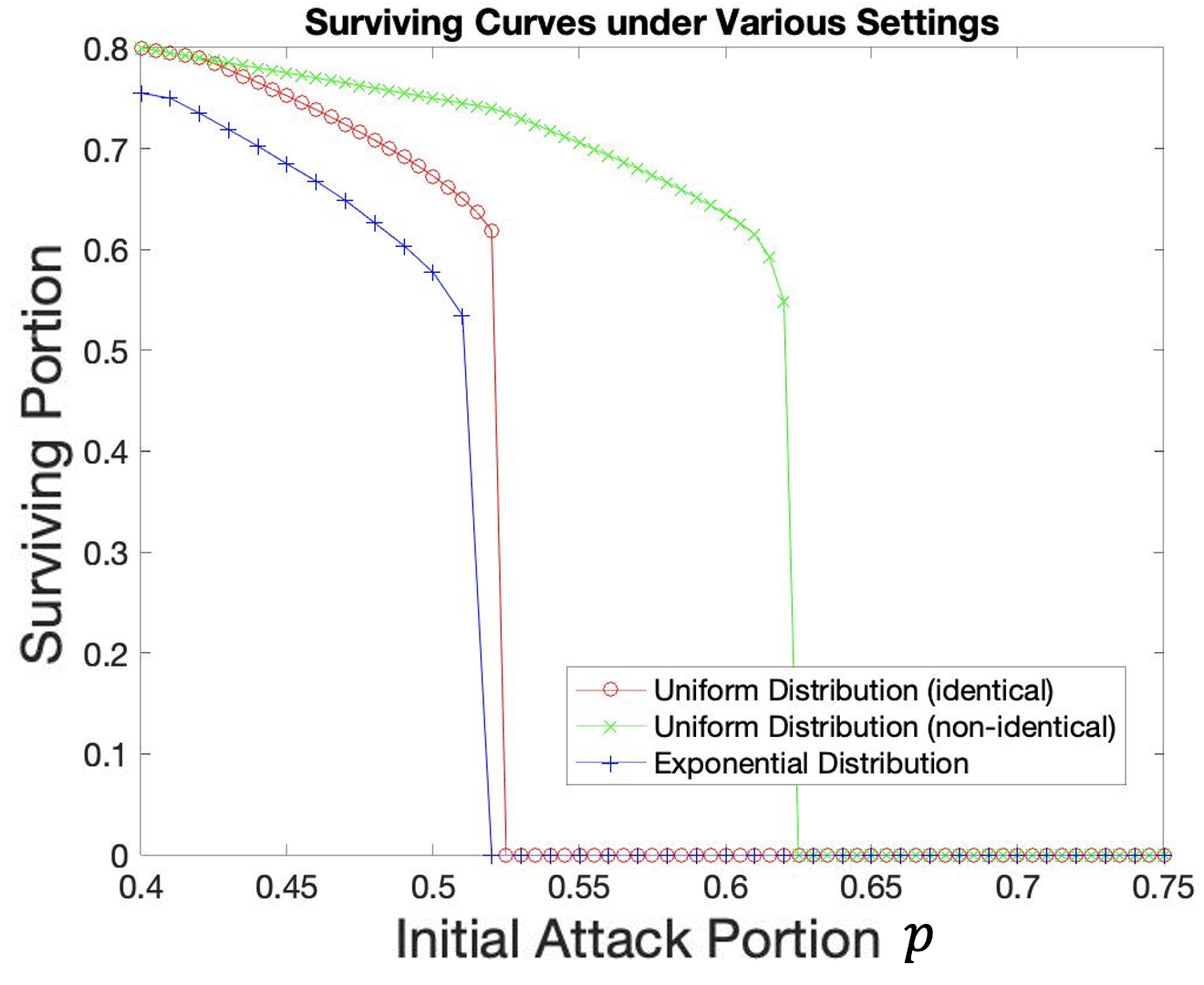}
    \caption{Surviving portion of the whole system under different system settings. The curve itself indicate the numerical results from the analytical equations while the points indicate the simulation results. We can see that the two results perfectly matched.}
    \label{fig:SCTN}
\end{figure}

The results are shown in Figure \ref{fig:SCTN}. The curve is a plot of numerical results from the analytical equations (6)-(14) and the points are the results from simulation. The x-axis of the plot indicates the different initial attack sizes. In this plot, the initial attack only occurred at network A. The y-axis of the plot indicates the surviving portion of the whole system, i.e., the sum of surviving nodes of network A and network B divided by the sum of initial number of nodes in network A and network B (2 million nodes).

The specific settings used in the three different curves in the order of red, green, and blue are:

\begin{itemize}
    \item \textbf{Uniform Distribution (identical): } $S_A, S_B \sim U(20, 180)$, $\mathbb{E}[L_A] = \mathbb{E}[L_B] = 75$. 
    \item \textbf{Uniform Distribution (non-identical): } $S_A \sim U(20, 180)$, $S_B \sim U(40, 280)$, $\mathbb{E}[L_A] = \mathbb{E}[L_B] = 75$.
    \item \textbf{Exponential Distribution: } $S_A, S_B \sim 20 + Exponential(\frac{1}{120})$, $\mathbb{E}[L_A] = \mathbb{E}[L_B] = 60$. 
\end{itemize}

From the figure, the analytical results match the simulation results under all different settings. We further observe that though the average free space in the third (Exponential distribution) case is set to be greater than the first (uniform distribution (identical)) case and the initial average load in the third case is set to be less than the first case, the third case still has less robustness (i.e. smaller surviving portion under the same initial attack size) compared to the first case. 

Initially the identical case of both uniform distribution and exponential distribution 
have the same initial attack size and same extra load to be redistributed in step 1. If they take the same coupling coefficients initially, the average extra load will be the same for both cases during the first time step. Since the second case has more nodes having less free space due to the shape of the exponential distribution, under the same average extra load, the second case will have more nodes fail during the first time step, which causes more extra load to be redistributed in the following time step. These additional loads add up over time, making the second case accumulate much more extra load compared to the first case, thus having a smaller final system size and being more vulnerable to the cascading failure.

\subsection{Non-identical Networks}

\begin{figure}[!h]
    \centering
    \includegraphics[width=1.0\linewidth]{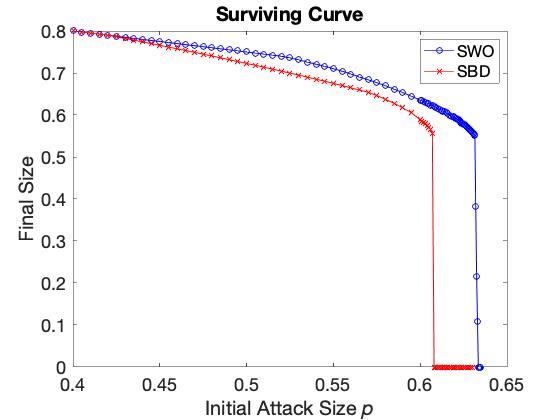}
    \caption{Comparison between the proposed SWO method for non-identical networks case and SBD method which degenerate to the case of a single complete graph. For both network, the expected load is set to be $L=75$ and the free space distribution are $S_A ~ U(20, 180)$ and $S_B ~ U(40,280)$ respectively. The results shows that the SWO strategy for general case is more robust in terms of both remaining portion under same initial attack size and the critical attack size.}
    \label{fig:HET}
\end{figure}

Previously, we claimed that our SWO strategy outperforms the SBD strategy and the FCC strategy under non-identical network settings. We now verify this claim. We show the results of the experiments using the SWO strategy we proposed and the SBD strategy we used previously. Here we set the average of the initial load for both networks $L=75$, and the free space distributions of networks A and B are $S_A \sim U(20, 180)$ and $S_B \sim U(40, 280)$ respectively. Figure \ref{fig:HET} illustrates the results under these settings. We can see that the proposed SWO strategy maintains its advantage over SBD in the non-identical case.

\begin{figure}[!h]
    \centering
    \includegraphics[width=1.0\linewidth]{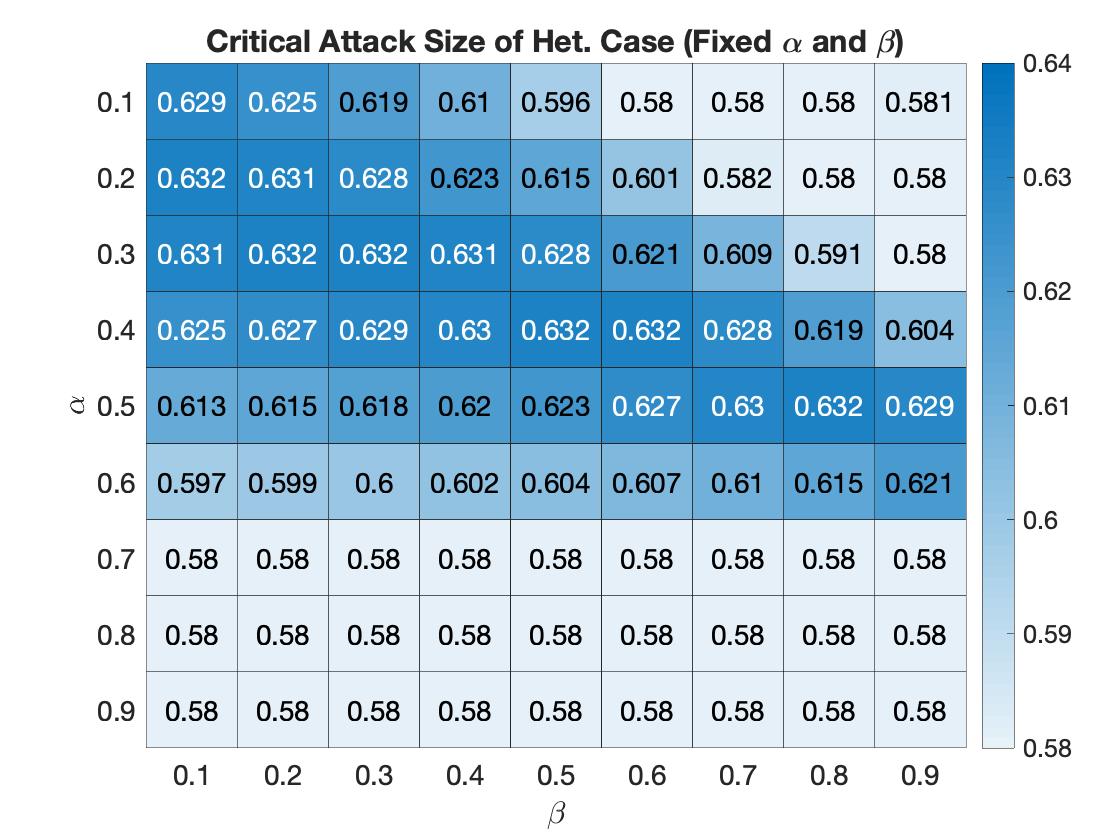}
    \caption{Heatmap for the critical attack size of different fixed coupling coefficients settings. The threshold is set to be 0.58, if the critical attack size is less than 0.58 it will shows 0.58 in the block.}
    \label{fig:FCC}
\end{figure}

If we compare the SWO strategy with the FCC strategy, the critical attack size of the SWO strategy is $0.634$, which is slightly higher than the critical attack size under the optimal settings of the FCC strategy (0.632), as shown in the heatmap in Figure \ref{fig:FCC}. 

\subsection{Local Redistribution According to Different Network Topologies}

Our methods and analysis were based on the fully-connected networks, but many real world networks do not have fully connected topologies. 
The network topology might influence the performance of such coupling systems and our proposed SWO method may not be as useful as before. In this section, we show that SWO is robust to different network topologies. 

We choose two different kinds of networks, which are the "Erdős–Rényi Random Graph" and the "Barabási–Albert Model". The first graph contains $N=100000$ nodes 
and the link between every pair of nodes exists with probability $p$. The second graph also contains $N=100000$ nodes and uses preferential attachment, where probability of an edge forming while generating the graph is proportional to the node degree. 
The final graph thus has degree distribution following the power law.

Since the network is not fully-connected anymore, any extra load is no longer globally redistributed. Instead, it is redistributed locally according to the network topology. If a node fails, the internally redistributed portion of its load is equally redistributed to its neighbors. To model external redistribution, we set the number of nodes inside both network A and network B to be the same, though they need not have the same network topology. 
We then index all nodes in each network. If two nodes in different networks have the same index number, we consider them to be connected. For example, in a transportation network, the different networks represent different transportation systems, and nodes with the same index number are at the same geographical location. The external distribution of a single node's extra load will be equally redistributed to its connected node and the neighbors of its connected node. If the node that fails is isolated, in other words, has no neighboring nodes, then the internal load will be redistributed to all the nodes in the same network. If both the node's connected node in the other network and the neighbors of that connected node are failed, the external load is also being equally redistributed to all nodes in the other network. This mechanism prevents the extra load from disappearing.

\begin{figure}[!h]
    \centering
    \includegraphics[width=1.0\linewidth]{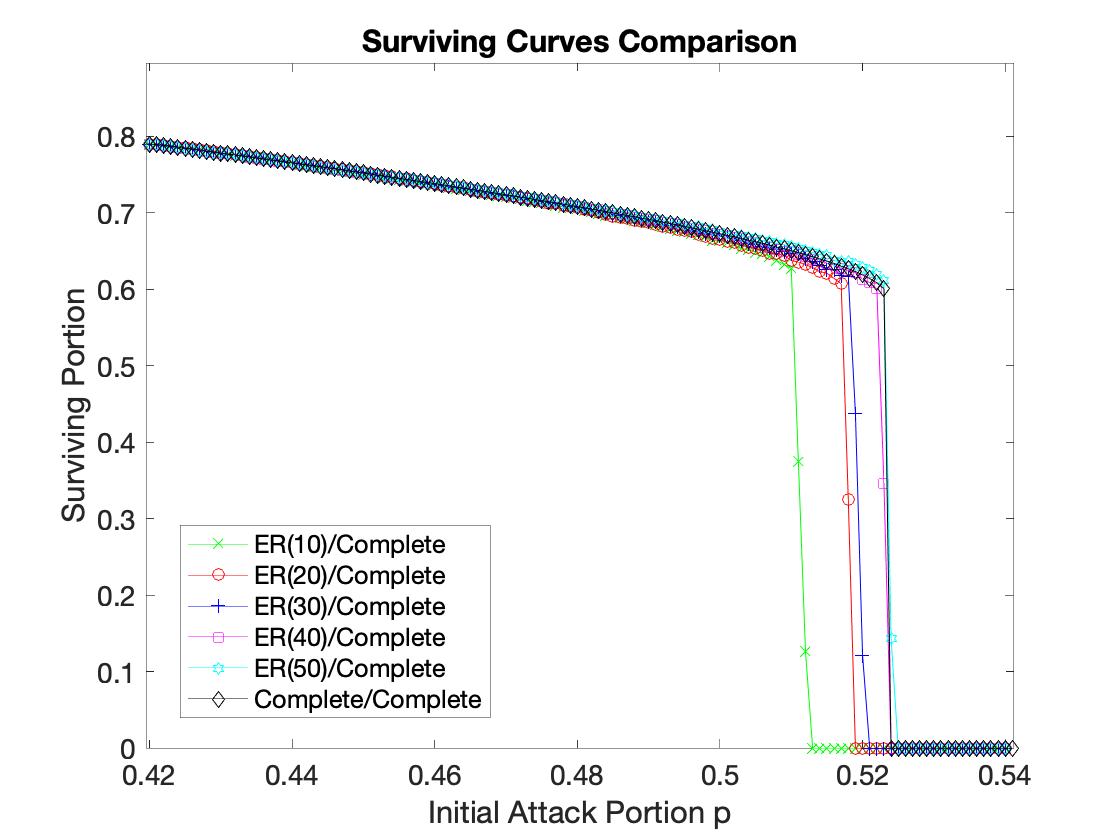}
    \caption{Comparison of ER-Random Graph with different average degree and the fully-connected (complete) graph case. The 4 colored curves are ER random graph with average degree ranging from 10 to 40 and the black curve indicates the fully-connected (complete) graph case. We can see that the higher average degree the ER random graph has, the better robustness of the system is, which can endure higher initial attack portion.}
    \label{fig:ERC}
\end{figure}

Figure \ref{fig:ERC} shows the results of different networks under our SWO strategy. We average our results over 100 experiments. Network A is set to be a complete graph while network B is an ER random graph.
From the results, we can see that SWO works well when one of the networks follows local redistribution according to its network topology. When the ER random graph has average degree equal to 40, the results are very close to the fully-connected case.

\begin{figure}[t]
    \centering
    \includegraphics[width=1.0\linewidth]{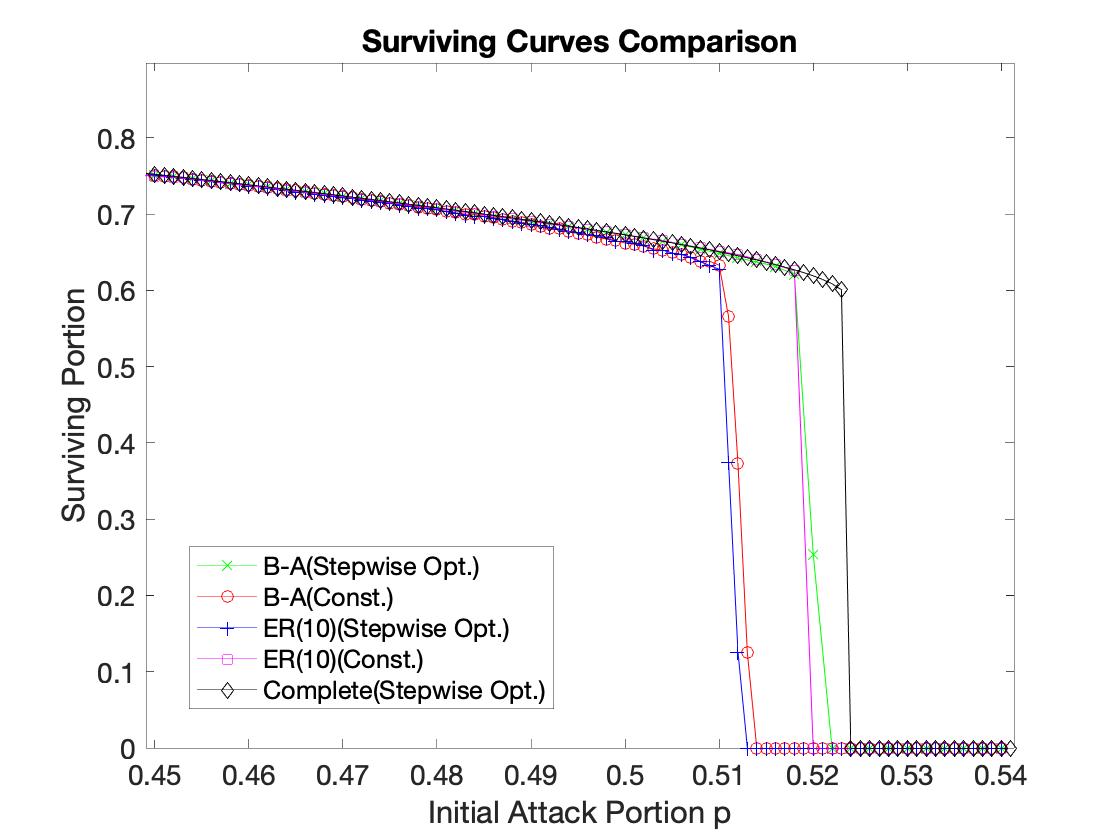}
    \caption{Comparison of ER-Random Graph and B-A Model using different strategies,  and the fully-connected (complete) graph case using step-wise optimization strategy. The results shows that the step-wise optimization (SWO) strategy outperforms the fixed strategies (FCC) in ER-Random graph.}
    \label{fig:BAERCO}
\end{figure}

If we set network B to be generated by the B-A instead of ER model, the results in Figure \ref{fig:BAERCO} shows that SWO's performance is much worse than that of the ER random graph under the same average degree. Since the degree variance is large in B-A model, nodes with large degree will likely receive more extra load and fail easily even when the attack size is small, thus triggering the cascading failure process. We discuss this result further in Section \ref{sec:discussion}. 

\begin{figure*}
\begin{subfigure}{0.32\textwidth}
    \centering
    \includegraphics[width=1.0\linewidth]{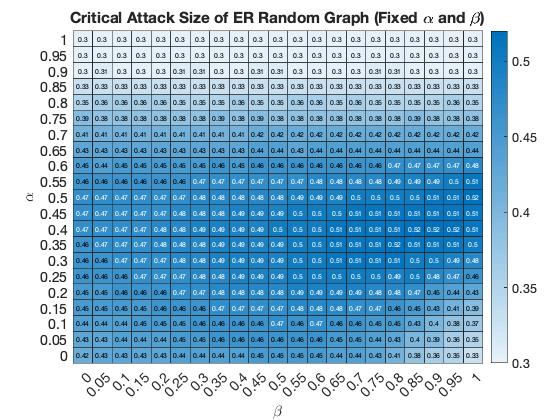}
    \caption{Heat map of the critical attack size for ER random graph under different coupling coefficients. The threshold is set to be 0.3, if the critical attack size is less than 0.3 it will shows 0.3 in the block.}
    \label{fig:ERFCC}
\end{subfigure}
\hfill
\begin{subfigure}{0.32\textwidth}
    \centering
    \includegraphics[width=1.0\linewidth]{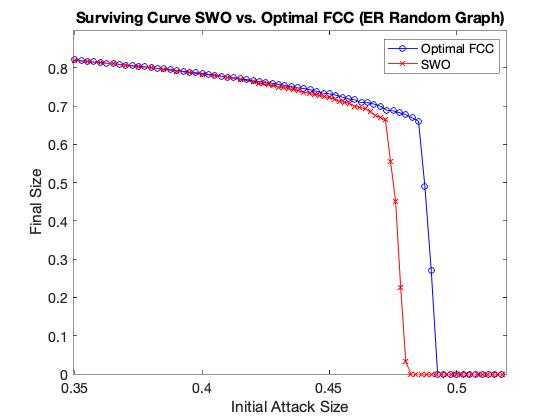}
    \caption{Surviving Curve of the SWO strategy and the optimal settings of FCC strategy.}
    \label{fig:ERComOpt}
\end{subfigure}
\hfill
\begin{subfigure}{0.32\textwidth}
    \centering
    \includegraphics[width=1.0\linewidth]{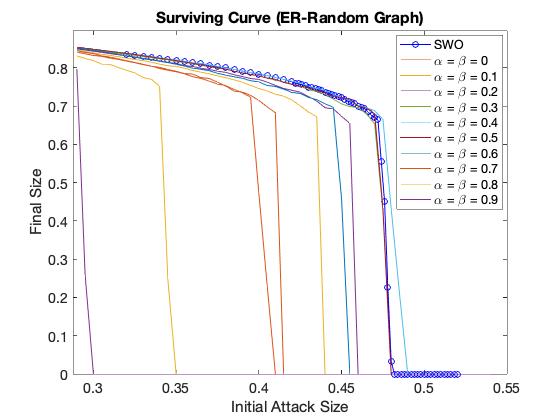}
    \caption{Surviving Curve of the SWO strategy and the different settings of FCC strategy when $\alpha = \beta$.}
    \label{fig:ERComSym}
\end{subfigure}
\caption{Results for ER-Random Graph}
\label{fig:ERR}
\end{figure*}

Next, we compare the FCC strategies with the SWO strategy we proposed. We start with ER random graph. The two networks in the system are both ER random graphs, with network A having average degree 20 and network B average degree 40. Before comparing the two different strategies, we find the performance of FCC strategies under different settings. 
To find the performance under different choices of fixed coupling coefficients and compute the best settings of $\{ \alpha, \beta \}$, we discretize the possible coupling coefficient values into 21 different values starting from 0, increasing by 0.05 and ending at 1. For the 2-network system, there are thus 21 times 21, totaling 441, different pairs of the coupling coefficients. We use brute force search to find out the critical attack size of each pair of the coupling coefficients. Figure \ref{fig:ERFCC} shows the heat map of the critical attack size under different coupling coefficients. The x-axis is the value of $\alpha$ while the y-axis is the value of $\beta$. The color and the number in each block indicates the critical attack size.

From the results in Figure \ref{fig:ERFCC}, we can see that the critical attack size mostly lies between 0.3 and 0.5. The optimal coupling coefficients are around $\{ \alpha, \beta \} = \{ 0.4, 0.9 \}$, for which the critical attack size is around 0.52.

Now we turn our attention to the performance of our SWO strategy. In our simulation, the critical attack size of the SWO strategy in this case is around 0.49. We plot the surviving curve of the SWO strategy together with the best FCC strategy where $\{ \alpha, \beta \} = \{ 0.4, 0.9 \}$ in Figure \ref{fig:ERComOpt}. We can see that in terms of critical attack size, our SWO strategy performs slightly worse than the best FCC strategy. However, in terms of final system size under certain initial attack size, our SWO strategy has similar performance to that of the FCC strategy before the system failed.

Our choice of fixed coupling coefficients depends on the fact that the initial failure only happens in network A. If the initial failure happens in network B, setting the coupling coefficients $\{ \alpha, \beta \} = \{ 0.4, 0.9 \}$ is equivalent to setting $\{ \alpha, \beta \} = \{ 0.9, 0.4 \}$ when the initial failure happens in network A. The critical attack size would then decrease to 0.3 according to Figure \ref{fig:ERFCC}.

In practice, we may not be able to anticipate where the initial attack happens in time to choose the right coupling coefficients. Thus, we would like our method to be robust to different attacks. A safer way to pick the pair of coupling coefficients is to let the pair of $\{ \alpha, \beta \}$ be symmetric, that is, $\alpha = \beta$. Under this case, if the initial attack takes place in the other network, the process of cascading failure may not change drastically like the previous example. We show the comparison of our SWO strategy and different FCC strategies with $\alpha = \beta$ in Figure \ref{fig:ERComSym}. The results show that our SWO strategy outperforms most of the settings of FCC, only being slightly worse than the settings that $\{ \alpha, \beta \} = \{ 0.4, 0.4 \}$. However, the initial failure may happen in either network. No matter what's the distribution of the initial attack among two networks, our SWO strategy could take this into account and react accordingly, while the FCC with preset coupling coefficients may not adapt to different kinds of attack. 

\begin{figure*}
\begin{subfigure}{0.32\textwidth}
    \centering
    \includegraphics[width=1.0\linewidth]{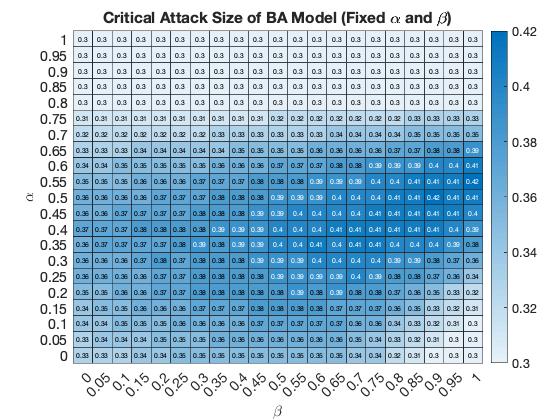}
    \caption{Heat map of the critical attack size for BA model under different coupling coefficients. The threshold is set to be 0.3, if the critical attack size is less than 0.3 it will shows 0.3 in the block.}
    \label{fig:BAFCC}
\end{subfigure}
\hfill
\begin{subfigure}{0.32\textwidth}
    \centering
    \includegraphics[width=1.0\linewidth]{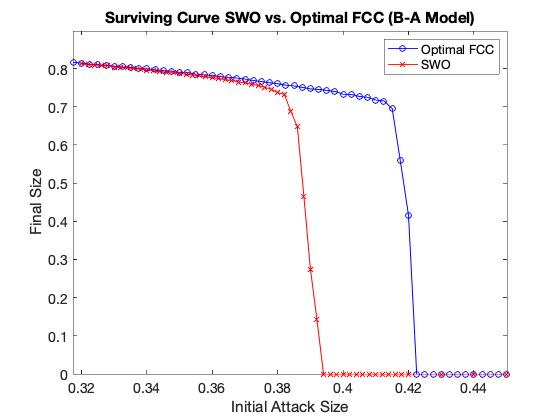}
    \caption{Surviving Curve of the SWO strategy and the optimal settings of FCC strategy.}
    \label{fig:BAComOpt}
\end{subfigure}
\hfill
\begin{subfigure}{0.32\textwidth}
    \centering
    \includegraphics[width=1.0\linewidth]{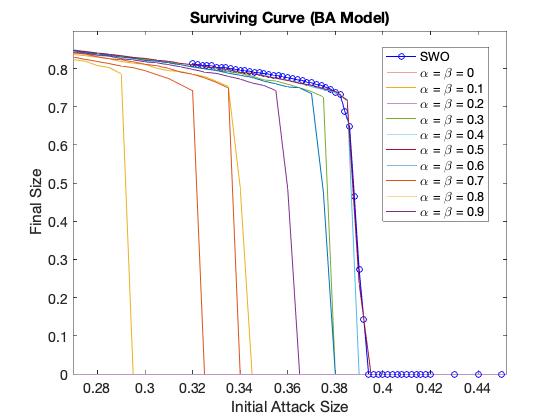}
    \caption{Surviving Curve of the SWO strategy and the different settings of FCC strategy when $\alpha = \beta$.}
    \label{fig:BAComSym}
\end{subfigure}
\caption{Results for B-A Model}
\label{fig:BAR}
\end{figure*}

Next, we run a series of experiments on networks with the B-A model. The two networks are both generated by the B-A model. Network A has average degree 20, and network B has average degree 40. Again, we find the performance of the FCC strategies under different settings by constructing the heat map of the FCC strategies for this system. Figure \ref{fig:BAFCC} shows the heat map of the critical attack size under different coupling coefficients.

From the results in Figure \ref{fig:BAFCC} we can see that the critical attack size mostly lies between 0.3 and 0.4. The optimal settings of the coupling coefficients are around $\{ \alpha, \beta \} = \{ 0.5, 0.9 \}$, for which the critical attack size is around 0.42.

Though the networks in our experiments with ER random graph and B-A model have the same average degree settings for the networks, the B-A model has much less robustness compared to the ER random graph. This is due to the great variance of the degree of the nodes in the B-A network. The nodes with large degree will then likely receive lots of extra loads and fail in the first few time steps once the cascading failure process starts.

We now compare the performance of our SWO strategy and the FCC strategy under the B-A model. In our simulation, the critical attack size of the SWO strategy is around 0.396. We plot the surviving curve of the SWO strategy together with the best FCC strategy where $\{ \alpha, \beta \} = \{ 0.5, 0.9 \}$ in Figure \ref{fig:BAComOpt}. We can see that in terms of critical attack size, our SWO strategy performs slightly worse than the best FCC strategy. However, in terms of final system size under a given initial attack size, our SWO strategy has similar performance at attacks smaller than the critical attack size. This result is similar to the case with the ER random graph.

Again, we assume the initial failure only happens in network A. If the initial failure instead happens in network B, the coupling coefficients settings of $\{ \alpha, \beta \} = \{ 0.5, 0.9 \}$ will lead to performance equivalent to setting $\{ \alpha, \beta \} = \{ 0.9, 0.5 \}$ when the initial failure happens only in network A, in which case the critical attack size decreases to 0.3 according to Figure \ref{fig:BAFCC}.

Similarly, we compare our SWO strategy and different FCC strategies with $\alpha = \beta$ in Figure \ref{fig:BAComSym}. The results shows that our SWO strategy outperforms every settings of FCC, have equal performance as the settings that $\{ \alpha, \beta \} = \{ 0.5, 0.5 \}$. Again, our SWO strategy could adapt to different initial attack and react accordingly while the FCC with preset coupling could not. 

\section{Discussion}\label{sec:discussion}

In this section, we focus on three different topics. First, we show that SWO can be extended to a multiple network system. Second, we discuss the case of extra load redistribution with arbitrary network topology. Third, we summarize the advantages of our SWO strategy compared to previously proposed fixed coupling strategies.

\subsection{Extension to Multiple Network System}
Following the definition in Section 2.3, for a $n$-network system, the coupling matrix at time $t$ is written as $\mathcal{M} (t)$ with size $n \times n$, where the element $m_{i, j} (t)$ at the $i$-th row, $j$-th column of the matrix is the fraction of the extra load to be redistributed from network $i$ to network $j$. We can see that when the number of network $n$ grows, the number of coupling coefficients grows in the order of $\Theta(n^2)$. Extending Definition \ref{def: opt} to $n$ networks, the objective function of the step-wise optimization problem now becomes:

\begin{equation*}
   \min _{\mathcal{M} (t)}\left(\sum_{i \in \mathcal{X}}\mathbb{E}\left[L_{i}+Q_{i(t+1)}\right] \Delta N_{i(t+1)}\right) \\ 
\end{equation*}

subject to:

\begin{equation*}
        \text{for } i, j \in \mathcal{X}, 0 \leq m_{i, j} (t) \leq 1 \\
\end{equation*}
\begin{equation}
        \forall i \in \mathcal{X}, \sum_{j \in \mathcal{X}} \leq m_{i, j} (t) = 1 \\
\end{equation}

Let us follow the special case of uniform distribution of free space discussed in Section 3.4. From Definition \ref{def: optid}, we can write the above optimization problem as:

\begin{equation*}
         \min _{\mathcal{M} (t)} \left(\sum_{i \in \mathcal{X}}N_i (1-p_i) \cdot \delta_{i} \left[ \delta_{i} + \left( L - Q_{i(t-1)}\right)\right]\right)
\end{equation*}

Subject to:

\begin{equation*}
        \text{for } i, j \in \mathcal{X}, 0 \leq m_{i, j} (t) \leq 1 \\
\end{equation*}
\begin{equation}
        \forall i \in \mathcal{X}, \sum_{j \in \mathcal{X}} \leq m_{i, j} (t) = 1 \\
\end{equation}

Where:

\begin{equation}
        \delta_{i} = \sum_{j \in \mathcal{X}}\left( \frac{m_{i, j} (t)(L+Q_{j(t-1)})\Delta Q_{j(t-1)}(1-p_j)N_j}{(1-p_i)dN_i(S_0 + d - Q_{it})}\right) \\
\end{equation}

We observe that the variables $\delta_{i}$ for $i \in \mathcal{X}$ are a linear transformation of the original coupling coefficients $m_{i, j} (t)$, and thus the solution set after the transformation is also a convex set. If we use the $\delta_{i}$ as our optimization variable, we can see there are no cross terms in the objective function. This implies that the Hessian matrix of this optimization problem is a diagonal matrix with non-negative diagonal elements. Thus, the Hessian matrix is positive semi-definite and the optimization problem is a convex optimization (which can be easily solved with standard algorithms). 

Since the number of coupling coefficients grows in the order of $\Theta(n^2)$, using brute force to search for the optimal fixed coupling coefficients is not possible in a reasonable amount of time as the number of networks $n$ grows larger. This scalability shows the benefit of our step-wise optimization algorithms under uniform free space distributions, which require solving a convex optimization problem, compared to the fixed coupling coefficients strategy. 

\subsection{Arbitrary Network Topology}
In the results in Section 4.5 with network topologies that are not fully connected, we can see that in some cases, our proposed SWO algorithm might not work as well as the fully-connected case 
due to the wrong estimation of load / free space distribution. Here we discuss the performance of our proposed algorithms in two different terms: average degree of the network and the variance of degree of the network.

Networks with greater average node degree yield better robustness of the system in both our SWO method and FCC method. Intuitively, when a network system has low average degree, once a node fails, only a few nodes could help to share the extra load. For example, for a network with average degree 2, a node that fails will in expectation only share its extra load with two other nodes. It is then likely that these neighboring nodes will receive a large amount of extra load and fail immediately. Thus, in the very beginning of the cascading failure process, the sequential failure of the nodes might be initiated from the part of the network with low degree and high clustering, commencing a series of local failures that eventually spread to the whole network. This causes the network to have weak robustness against cascading failure, no matter which strategy we choose for coupling between networks. The results in Section 4.4 for the ER Random Graph and B-A Model under different average degrees illustrate this intuition.

\begin{figure}[!h]
    \centering
    \includegraphics[width=1.0\linewidth]{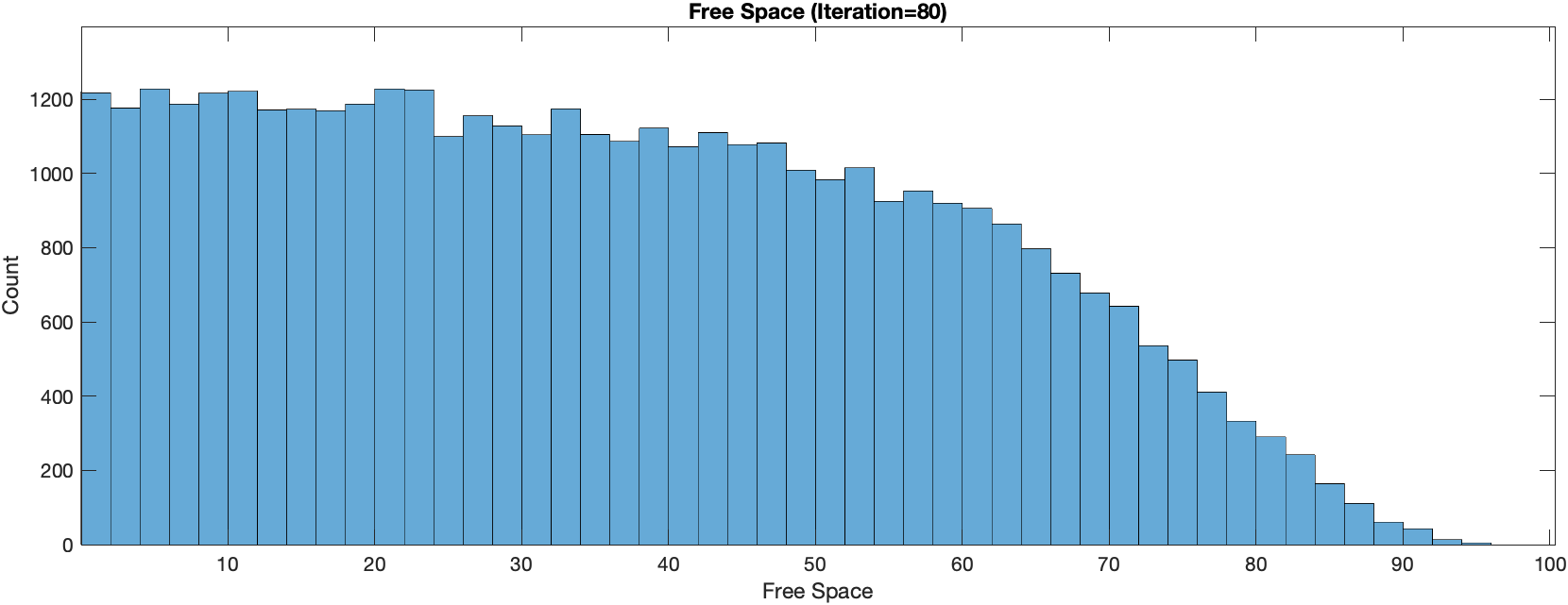}
    \caption{Free space distribution of B-A model running the step-wise optimization strategy after 80 timesteps. We can see the distribution of the free space became distorted compare to the original uniform distribution.}
    \label{fig:80}
\end{figure}

Our proposed SWO algorithms might be more affected by a high variance of average degree
compare to the fixed coupling coefficients strategy, and thus more likely to perform worse. A greater variance of degree means that certain portions of the network might have low average degree, while other parts of the network might have higher average degree. 
No matter which coupling strategy we choose, the low average degree part is always a weakness of the system. However, since our proposed SWO algorithms assume a complete graph case when estimating the distribution of load and free space, greater variance of degree will have more impact on their performance. Since the distribution of the load and the free space might be distorted when the cascading failure process continues, our proposed algorithm might incorrectly estimate the distributions. Thus, when the time step increases, the coupling coefficients chosen might be farther from the exact optimal point. This intuition further explains why our SWO algorithm performs a little bit worse in the case of a B-A model compared to E-R random graph, 
since the variance of degree is higher for B-A models. Figure \ref{fig:80} illustrates this concept with the free space distribution of a B-A model graph after 80 timesteps of the cascading failure process. 
Our current SWO assumes the free space distribution to be uniform. A better estimation of the load / free space distribution is one future direction of this work, which will likely improve the performance of SWO when the network has high node degree variance.

\subsection{The Advantage of SWO}

\subsubsection{Computation Complexity}
The computation time required for our SWO algorithms is much less than that of finding the optimal fixed coupling coefficients, since (to the best of our knowledge) there is no efficient way to determine the optimal fixed coupling coefficients besides brute force search. Depending on the precision of the coupling coefficients and the initial attack size, if we wish to find coupling coefficients to a precision of $m_C$ digits, we need to run a total of $10^{2m_C}$ experiments for a two network system and $10^{2C^M_2m_C}$ experiments for $M$ networks system. If we want to measure the initial attack size to a precision of $m_A$ digits, we need to run $10^{m_A}$ different simulations. Moreover, if we want the results to be more accurate, we might average our results over $N_E$ experiments. In short, we need to run a total of $N_E \cdot 10^{2C^M_2m_C+m_A}$ experiments to find the optimal coupling coefficients. In our experiments of the two networks system with $10^5$ nodes in each network, it takes around $10.28$ seconds to run a single experiment for the fixed coupling coefficients setting, while it takes $19.35$ seconds to run a single experiment for our dynamic SWO strategy. If we want the precision of both the coupling coefficients and initial attack size to be 3 digits and take an average over $10$ experiments, we need to run $10^{10}$ experiments, which requires $5 \times 10^9$ times the amount of time to find the optimal coupling coefficients compared to our dynamic SWO strategy. 

\subsubsection{Scalability}

If the number of networks $M$ 
in the system increases, following the discussion in the previous Section 5.3.1, the required time to find the optimal settings of the coupling coefficients for FCC will increase exponentially with $M$. Our SWO method, however, still solves a convex optimization, more specifically, a quadratic convex optimization problem with number of variables increasing in the order of $\Theta (M^2)$ (the coupling coefficients) when the free space distribution is uniform or exponential. This process requires much shorter computation time to determine the coupling coefficients during the cascading failure process, compared to optimizing a fixed strategy. 
Uniform distribution can be a good approximation of a more homogeneous system where all nodes have its ability to deal with the work load (capacity) in a certain range according to the system specification. On the other hands, exponential distribution is a good approximation of the system where the capacity of a node, as well as the degree, follows a power law.

\subsubsection{Adaptability and Robustness}

As seen in Sections 4.1 and 4.2, our SWO strategy outperforms or matches the performance of all other methods for the identical and non-identical fully-connected network system. Also, our SWO strategy outperforms most FCC settings under the ER random graph and B-A model network topologies. If the networks are not identical, the SBD method will become worse than our SWO method and SWO still outperform most FCC settings and have the same performance as the optimal FCC settings. Varying the initial attack pattern (e.g., if we do not anticipate it correctly), can significantly change the optimal settings of FCC, while our SWO method could adapt to the actual observed attack. If there is a second wave of attack during the cascading failure process, for example, the optimal settings of FCC might change, while our SWO method could respond based on the current system situation. Our SWO method thus exhibits more flexibility compared to fixed baselines.

\section{Conclusion}\label{sec:conclusion}

In this work, we studied the robustness of different coupling strategies of interdependent networks based on a flow-redistribution model. We proposed a step-wise optimization strategy to dynamically adjust the coupling coefficients between networks according to the current system situation. Comparing to the fixed coupling coefficients strategy, to our best knowledge, there is not a clever way to find the optimal settings of the coupling coefficients except brute force search, but our method exhibits much better computational complexity. Our strategy provides a clear framework based on minimizing the subsequent total extra load in each time step, which is proved to be robust against cascading failures in several different situations. We admit higher critical attack sizes than fixed coupling coefficients strategies, and our proposed method approaches the performance of the SBD strategy under the identical networks setting, which is known to exhibit a good robustness against cascading failure. Our work provides a framework for dynamically searching for the optimal coupling coefficients and opens up many directions for future works. An immediate direction will be searching for an optimal strategy for setting the coupling coefficients under different network topologies. As discussed in Section \ref{sec:discussion}, a high variance in degree distribution may distort the free space distribution in later time steps, making SWO's estimation of the free space distribution more and more inaccurate over time. Combining an intelligent strategy for setting the coupling coefficients with carefully chosen strategies for local node redistribution may also alleviate errors in estimating the free space distribution. 
Searching for an optimal combination of local redistribution strategies and dynamic coupling strategies is an exciting direction for future work. Our work on dynamic coupling strategies could act as a basis for these future topics.

\ifCLASSOPTIONcompsoc
  \section*{Acknowledgments}
\else
  \section*{Acknowledgment}
\fi

This work was supported in part by the National University Transportation Center for Improving the Mobility of People and Goods (Mobility21) and the Scott Institute for Energy Innovation.

\ifCLASSOPTIONcaptionsoff
  \newpage
\fi



%
\bibliographystyle{ieeetr}
\bibliography{ref}




%

\begin{IEEEbiography}[{\includegraphics[width=1in,height=1.25in,clip,keepaspectratio]{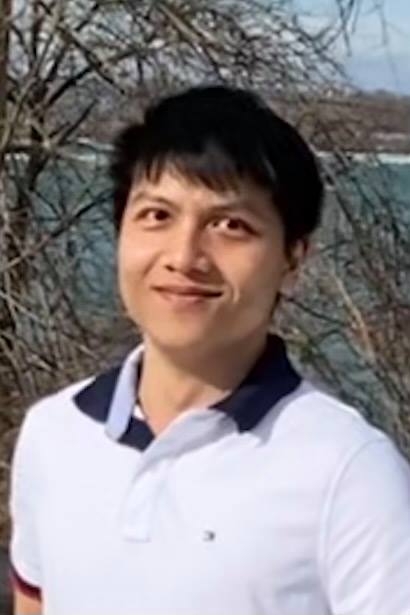}}]{I-Cheng Lin}
received the B.S. degree in Electrical Engineering from National Taiwan University, Taipei, Taiwan, in 2015, and the M.S. degree in communication engineering from the Graduate Institute
of Communication Engineering, National Taiwan University in 2017.  He is currently pursuing the Ph.D. degree in Electrical and Computer Engineering in Carnegie Mellon University, Pittsburgh, PA. His research interests including analysis and optimization of networked systems, with emphasis on applications of machine learning and resilience of networked systems such as intelligent transportation systems.
\end{IEEEbiography}

\begin{IEEEbiography}[{\includegraphics[width=1in,height=1.25in,clip,keepaspectratio]{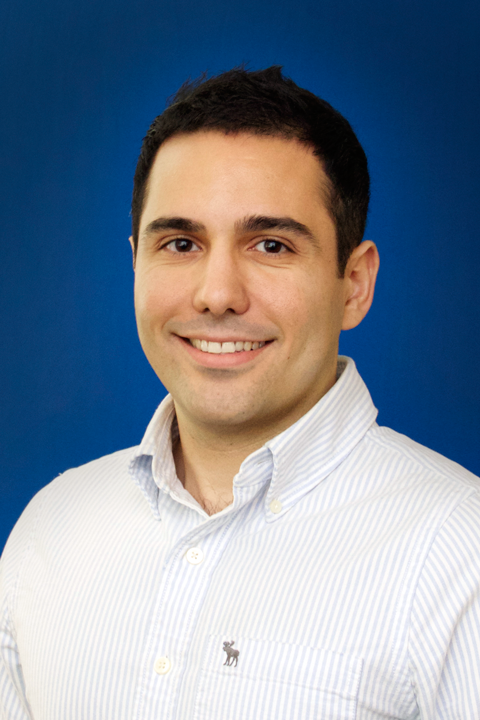}}]{Osman Ya\u{g}an}
received the B.S. degree in Electrical and Electronics Engineering from Middle East Technical University, Ankara (Turkey) in 2007, and the Ph.D. degree in Electrical and Computer Engineering from University of Maryland, College Park, MD in 2011. In August 2013, he joined the faculty of the Department of Electrical and Computer Engineering at Carnegie Mellon University, where he is currently a  Research Professor. Dr. Yagan’s research is on modeling, analysis, and performance optimization of computing systems, and uses tools from applied probability, data science, machine learning, and network science. Specific topics include wireless communications, security, random graphs, social and information networks, and cyber-physical systems. He is a recipient of the CIT Dean’s Early Career Fellowship, IBM Faculty Award, and ICC 2021 Best Paper Award. \end{IEEEbiography}


\begin{IEEEbiography}[{\includegraphics[width=1in,height=1.25in,clip,keepaspectratio]{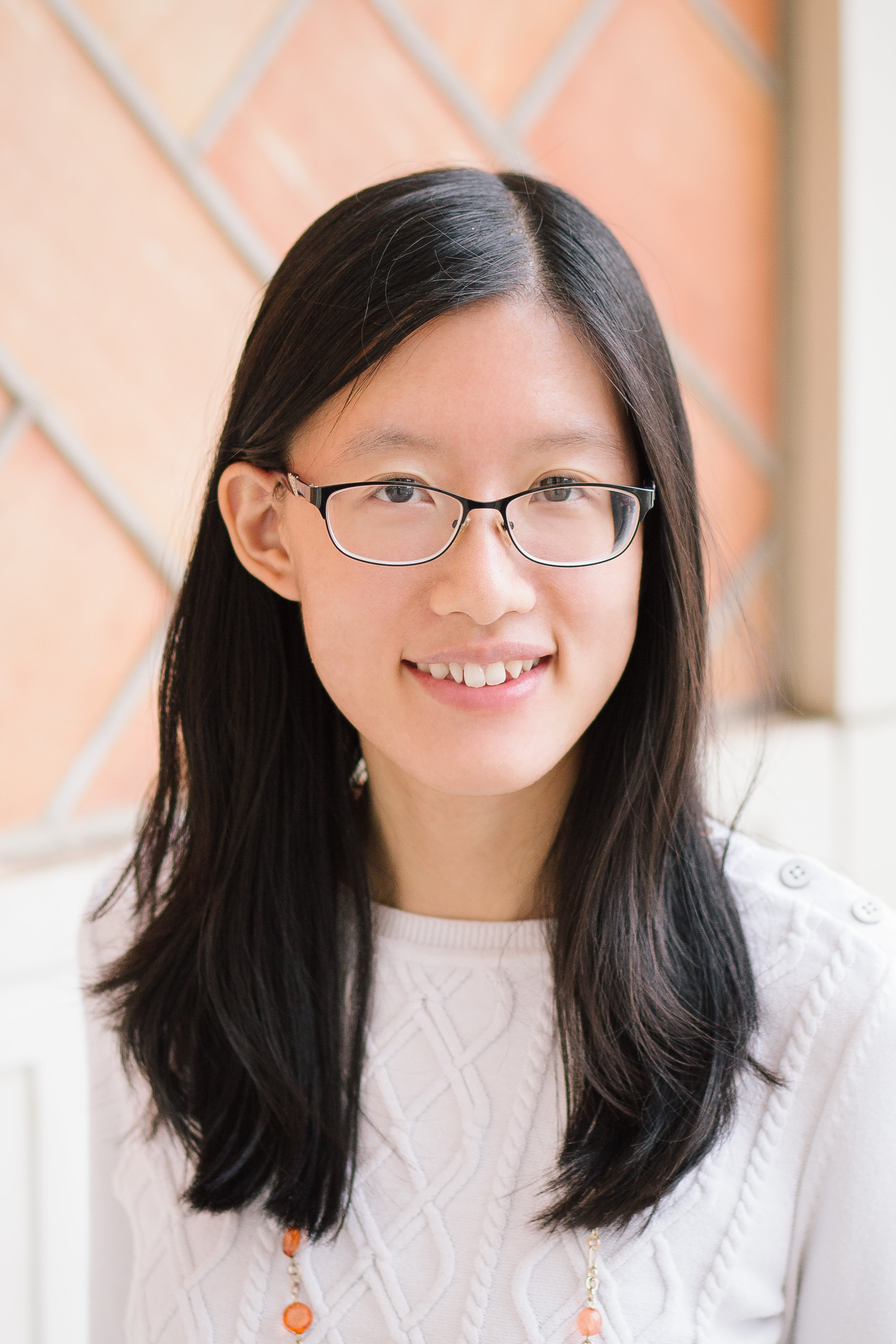}}]{Carlee Joe-Wong} is the Robert E. Doherty Associate Professor of Electrical and Computer Engineering at Carnegie Mellon University. She received her A.B. degree (\emph{magna cum laude}) in Mathematics, and M.A. and Ph.D. degrees in Applied and Computational Mathematics, from Princeton University in 2011, 2013, and 2016, respectively. Her research interests lie in optimizing various types of networked systems, including applications of machine learning and pricing to cloud computing, mobile/wireless networks, and transportation networks. From 2013 to 2014, she was the Director of Advanced Research at DataMi, a startup she co-founded from her research on mobile data pricing. She received the NSF CAREER award in 2018 and the ARO Young Investigator award in 2019.
\end{IEEEbiography}




\end{document}